\documentclass[iop]{emulateapj}

\usepackage{amsmath}
\usepackage{graphicx,epsfig,epsf,graphics,epstopdf}
\usepackage{natbib,hyperref}
\def\apj{ApJ}
\def\nat{Nature}
\def\mnras{MNRAS}
\def\apjl{ApJL}
\def\apjs{ApJS}

\def\aap{A\&A}
\def\aj{AJ}

\def\sun{\odot}

\def\teff{T_\mathrm{eff}}
\def\tsurf{T_\mathrm{surf}}

\shorttitle{UV Radiation on Earth-like Planets Orbiting M dwarfs}
\shortauthors{Rugheimer et al.}

\begin{document}

\title{Effect of UV Radiation on the Spectral Fingerprints of Earth-like Planets Orbiting M dwarfs}

\author{S. Rugheimer\altaffilmark{1,2}, L. Kaltenegger\altaffilmark{2,3}, A. Segura\altaffilmark{4}, J. Linsky\altaffilmark{5}, and S. Mohanty\altaffilmark{6}}

\submitted{Submitted August 18th, 2014. Accepted to ApJ June 10, 2015}

\altaffiltext{1}{Harvard Smithsonian Center for Astrophysics, 60 Garden st., 02138 MA Cambridge, USA}
\altaffiltext{2}{Carl Sagan Institute, Cornell University, Ithaca, NY 14853}
\altaffiltext{3}{MPIA, Koenigstuhl 17, 69117 Heidelberg, Germany}
\altaffiltext{4}{Instituto de Ciencias Nucleares, Universidad Nacional Aut\'onoma de M\'exico, M\'exico}
\altaffiltext{5}{JILA, University of Colorado and NIST, 440 UCB Boulder, CO 80309-0440, USA}
\altaffiltext{6}{Imperial College London, 1010 Blackett Lab, Prince Consort Rd., London SW7 2AZ,  UK}

\begin{abstract}
We model the atmospheres and spectra of Earth-like planets orbiting the entire grid of M dwarfs for active and inactive stellar models with $\teff$ = 2300K to $\teff$ = 3800K and for six observed MUSCLES M dwarfs with UV radiation data. We set the Earth-like planets at the 1AU equivalent distance and show spectra from the VIS to IR (0.4$\mu$m - 20$\mu$m) to compare detectability of features in different wavelength ranges with JWST and other future ground- and spaced-based missions to characterize exo-Earths. We focus on the effect of UV activity levels on detectable atmospheric features that indicate habitability on Earth, namely: H$_2$O, O$_3$, CH$_4$, N$_2$O and CH$_3$Cl. 

To observe signatures of life - O$_2$/O$_3$ in combination with reducing species like CH$_4$, we find that early and active M dwarfs are the best targets of the M star grid for future telescopes. The O$_2$ spectral feature at 0.76$\mu$m is increasingly difficult to detect in reflected light of later M dwarfs due to low stellar flux in that wavelength region. N$_2$O, another biosignature detectable in the IR, builds up to observable concentrations in our planetary models around M dwarfs with low UV flux. CH$_3$Cl could become detectable, depending on the depth of the overlapping N$_2$O feature. 

We present a spectral database of Earth-like planets around cool stars for directly imaged planets as a framework for interpreting future lightcurves, direct imaging, and secondary eclipse measurements of the atmospheres of terrestrial planets in the HZ to design and assess future telescope capabilities.

\end{abstract}


\section{INTRODUCTION}

About 2000 extrasolar planets have been found to date with thousands more awaiting confirmation from space and ground-based searches. Several of these planets have been found in or near the circumstellar Habitable Zone \citep[see e.g.][]{quintana2014, borucki2013, kaltenegger2013, batalha2013, borucki2011, kaltenegger2011, udry2007} with masses and radii consistent with rocky planet models. Future mission concepts to characterize Earth-like planets are designed to take spectra of extrasolar planets with the ultimate goal of remotely detecting atmospheric signatures \citep[e.g.][]{beichman1999, beichman2006, cash2006, traub2006, kaltenegger2006}. Several proposed missions are designed to characterize nearby Super-Earth and Earth-like planets using emergent visible and infrared spectra. For transiting terrestrial planets, the James Web Space Telescope \citep[JWST, see][]{gardner2006,deming2009,kaltenegger2009} as well as future ground and space based telescopes \citep{snellen2013, rodler2014} will search for biosignatures in a rocky planet's atmosphere. NASA's explorer mission, TESS, is designed to search the whole sky for potentially habitable planets around the closest and brightest stars to Earth \citep{ricker2014} for eventual follow-up with JWST and other large ground-based observatories such as the E-ELT or GMT. 

In our solar neighborhood, 75\% of stars are M dwarfs. The abundance of M dwarfs as well as the contrast ratio and transit probability favor the detection of planets in the Habitable Zone of M dwarfs. Therefore, it is likely that the first habitable planet suitable for follow-up observations will be found orbiting a nearby M dwarf \citep{dressing2013}. The M spectral class is very diverse, spanning nearly three orders of magnitude in luminosity and an order of magnitude in mass.

The UV environment of a host star dominates the photochemistry and therefore the resulting atmospheric constituents including biosignatures for terrestrial planets \citep[see e.g.][]{domagal2014, tian2014, grenfell2014, rugheimer2013, hu2012, segura2005}. To date, few observations exist in the UV region for M dwarfs. Previously, only UV spectra of very active flare stars, such as AD Leo, were available from the IUE satellite, primarily during flares and for a few quiescent phases. The MUSCLES program observed chromospheric emission from six weakly active M dwarfs with HST \citep{france2013}. Note that Ly-$\alpha$ is by far the strongest line in the UV for M dwarfs. Since the core of the intrinsic stellar Ly-$\alpha$ emission is absorbed by neutral hydrogen in the interstellar medium, one must reconstruct the line to compensate for this absorption to get accurate flux levels \citep[see e.g.][]{wood2005, linsky2013}. 

Several groups have explored the effect of a different stellar spectral type on the atmospheric composition of Earth-like planets by primarily considering one star, AD Leo, as a template for the diverse range of M dwarfs, and by using the extreme limit of inactivity, photosphere only, PHOENIX models \citep{grenfell2014, kitzmann2011a, kitzmann2011b, wordsworth2011, segura2005}. In this paper we expand on this work by establishing planetary atmosphere models for the full M dwarf main sequence, using a stellar temperature grid from 3800K to 2400K including recent HST observations of 6 M dwarfs to explore the effect of different spectral energy distributions on terrestrial atmosphere models and on detectable atmospheric signatures, including biosignatures. Atmospheric biosignatures are remotely detectable chemical species in the atmosphere that are byproducts of life processes. Recent research shows possible false positives may occur under certain geological settings \citep{domagal2014, tian2014, wordsworth2014}.

We create a grid of M dwarf input spectra, that can be used to probe the entire range of UV flux levels. To showcase the whole range, we simulate model planets around host stars at the two extreme limits of activity: active and inactive stellar models. We compare these models to the six observed M dwarfs with recent UV observations \citep{france2013}. 

We explore the influence of stellar UV flux on the atmospheric structure, chemical abundance, and spectral features for Earth-like planets including the observability of biosignatures in the VIS to IR. We focus our analysis on spectral biosignatures for a temperate rocky planet like Earth: O$_3$, O$_2$, CH$_4$, N$_2$O and CH$_3$Cl; and those that indicate habitability: H$_2$O and CO$_2$ \citep{desmarais2002,sagan1993,lovelock1975}. 

In \S2, we describe our model, and \S3 presents the influence of stellar types on the abundance of atmospheric chemical species. In \S4, we examine the remote observability of such spectral features, and in \S5 and \S6, we summarize the results and discussing their implications. 

\section{MODEL DESCRIPTION}

\subsection{Stellar M Dwarf Spectral Grid Model}

M dwarfs span nearly three orders of magnitude in luminosity and remain active for much longer timescales than earlier stellar types. As main sequence stars age, their UV flux levels decrease even as their bolometric luminosity increases. Understanding the UV flux incident on a planetary atmosphere is critical to understanding and interpreting future observations of atmospheric constituents, including biosignatures. 

Current stellar models are unable to model the UV region from M dwarfs self-consistently due to three main reasons. First, complex magnetic fields responsible for heating the chromosphere are thought to drive much of the UV activity and have thus far been ignored in stellar models. Second, the models are missing opacities in the UV, and third, the models are semi-empirical with no energy conservation to balance magnetic heating with radiative losses. These problems are being addressed by current work and models will be available in the future to test against M dwarf UV observations

\begin{figure}[h!]
\centering
\includegraphics[scale=0.37,angle=-90]{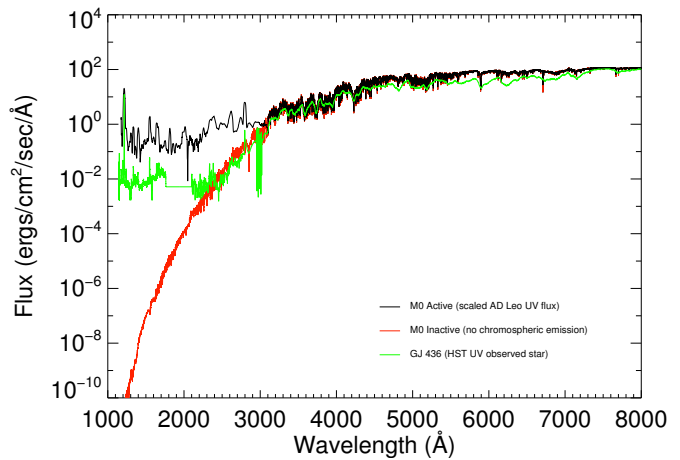}
\caption{Stellar input spectra at the top-of-atmosphere (TOA) of Earth-like planet at the 1AU equivalent of an M0 ``active'' star with UV flux scaled from AD Leo (black), ``inactive'' PHOENIX model of M0, and HST UV observations plus photosphere model for GJ436 (green).\label{photovschromo}}
\end{figure}

We generate stellar input models for the entire M dwarf spectral class (M0 to M9) to explore the boundaries of the UV environment of an exoplanet orbiting a main sequence star. We create two sets of models based on the extreme limits of stellar activity. For the purposes of this paper, ``active'' stellar models are constructed to represent the most active M dwarf measurements and ``inactive'' semi-empirical models without chromospheres are constructed to represent the lowest theoretical UV flux field. We compare these limiting-case models to six well-observed M dwarfs which show significant chromospheric flux (see Fig. \ref{photovschromo}) despite being traditionally classified as quiescent stars due to the presence of H$_\alpha$ in absorption \citep{france2013}. A few M dwarfs observed with GALEX in the NUV (1750\AA -- 2750\AA) show near photospheric continuum level fluxes (as computed with PHOENIX models) and can be classified as inactive even though the exact UV flux level has not been measured yet. For example, Gl 445 and Gl 682 have roughly 90\% of the observed near ultraviolet (NUV) flux density is predicted by PHOENIX model photospheres \citep{shkolnik2014}. More observations are needed to determine the ultimate lower limit of UV flux emitted by old and/or late M dwarfs. To compare our calculations for inactive star models to published work, we use theoretical photosphere PHOENIX models as a lower bound \citep{allard2014}.

\begin{figure}[h!]
\centering
\includegraphics[scale=0.35, angle=-90]{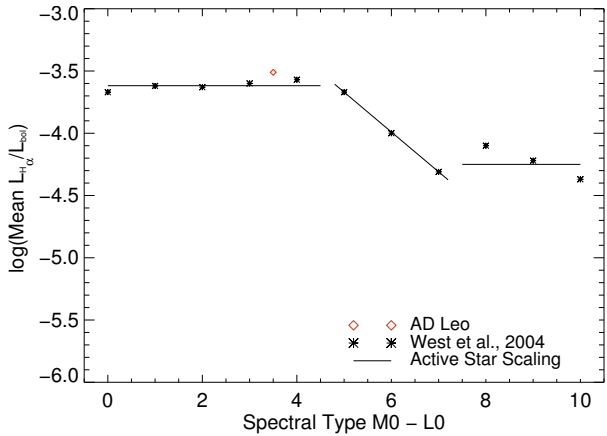}
\caption{Scaling used for active stellar models (black line) over plotted on \citet{west2004} measurements of log(mean $L_{H_{\alpha}}/L_{bol})$ (black asterisks).\label{scaling}}
\end{figure}

The chromospheric far-ultraviolet (FUV) flux from M, L and T dwarfs (i.e., very low mass stars and brown dwarfs), while crucial for determining the potential habitability of any planets around them, is very poorly characterized through direct observations. Given their common chromospheric origin, H$_\alpha$, Ly-$\alpha$, Ca II H and K, and Mg II H and K emission lines have all been used as proxies for FUV activity. However, many of these lines are inaccessible to M dwarfs because they are at shorter wavelengths where these  stars are intrinsically less luminous. For M dwarfs, H$_\alpha$ emission in particular has been studied extensively and to date has the most robust dataset for each M dwarf spectral type compared with the other lines considered above \citep{west2004, west2011}. In addition, \citet{jones2014} shows H$_\alpha$ fluxes correlate with NUV fluxes from GALEX. Therefore, we use H$_\alpha$ to estimate the FUV emission for our active stellar grid by scaling the known FUV and H$_\alpha$ emission from the active M3.5 star AD Leo (see  Eq. \ref{eqHalpha} and Fig. \ref{scaling}). Very active, early M dwarfs (M0-M5) are known to saturate in the H$_\alpha$ emission around log($L_{H_{\alpha}}/L_{bol})=-3.75$ \citep{hawley1996, reiners2008, west2004}.

\begin{figure*}[ht!]
\centering
\includegraphics[scale=0.57,angle=0]{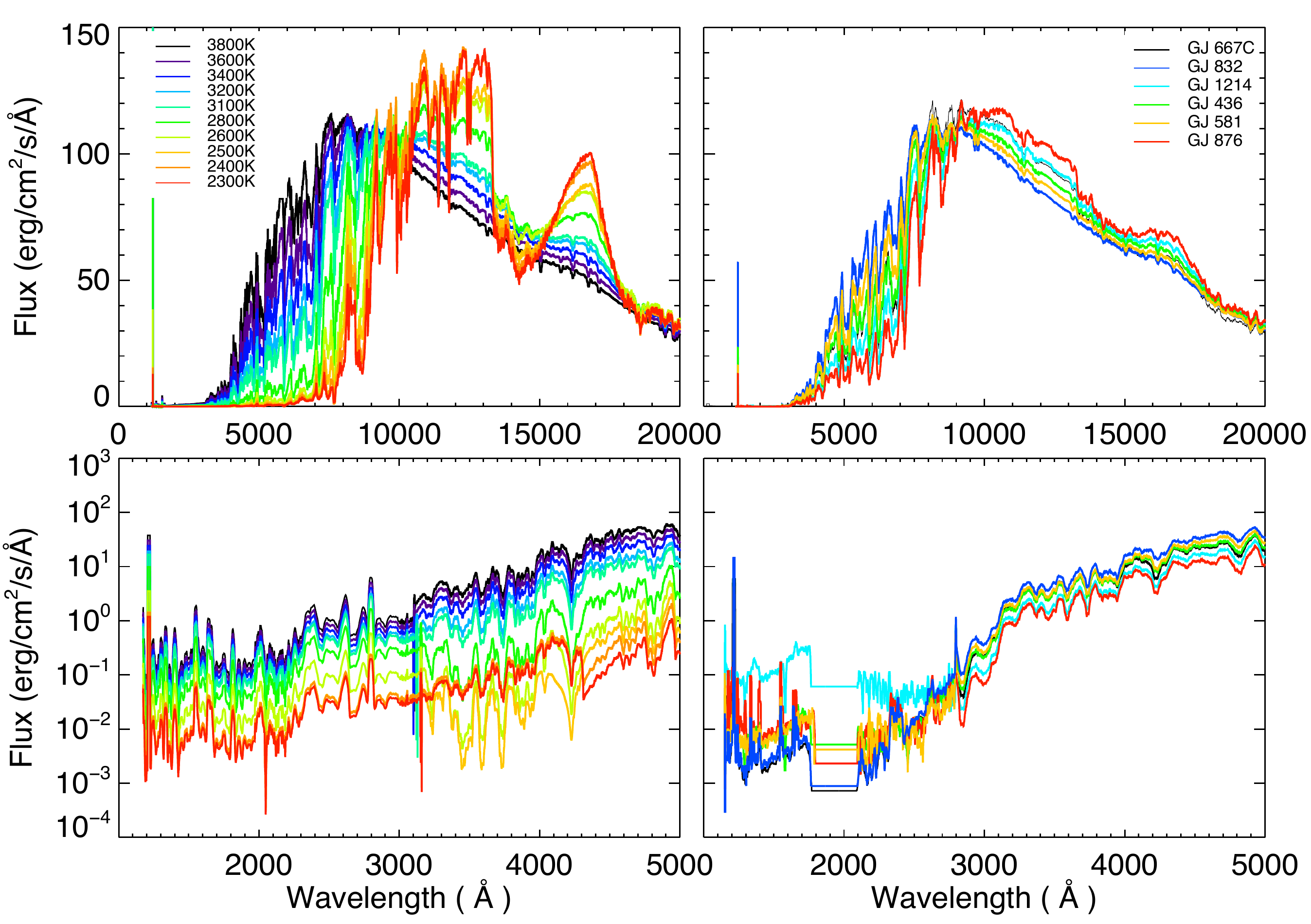}
\caption{Stellar input spectra (top) from 1000 to 20000 \AA\ for the M0-M9 active grid stars with UV scaled by H$_\alpha$ and AD Leo UV flux (left) and MUSCLES stars with HST UV observations (right). UV input stellar fluxes in log scale (bottom). \label{Mstellarspectra}}
\end{figure*}

\citet{west2004} tabulates the log(mean $L_{H_{\alpha}}/L_{bol}$) versus spectral type from M0-L0 from observations of 1910 active M dwarfs.  We parameterize the data from \citet{west2004} to derive a relationship of UV to mean H$_\alpha$ emission for each spectral sub type in the M dwarf class (See Fig. \ref{scaling} and Eq \ref{eqHalpha}).
\begin{eqnarray}\label{eqHalpha}
\textrm{log} (L_{H_\alpha *} / L_{bol *})  &=& -3.62 \text{ for M0 - M4}  \\
\textrm{log} (L_{H_\alpha *} / L_{bol *}) &= &-0.32\times(M) - 2.07, \nonumber \\
& &\text{where M} = {5,6,7} \text{ for M5, M6, M7}  \nonumber \\
\textrm{log} (L_{H_\alpha *} / L_{bol *}) &= &- 4.25 \text{ for M8 - M9}   \nonumber  
\end{eqnarray}
We use the H$_\alpha$/UV scaling of AD Leo, which has both a well characterized UV spectrum and H$_\alpha$ emission measurement as a calibration data point to set the scale. AD Leo, an M3.5eV star, has log($L_{H_\alpha} / L_{bol}$) = -3.51 \citep{walkowicz2009} and has been well observed in the UV (see \citet{segura2005} for AD Leo IUE spectrum and \citet{linsky2013} for reconstructed Ly-$\alpha$ flux). The M dwarf active spectrum is then given by:
\begin{eqnarray}\label{finalequation}
F_{FUV*} = F_{FUV\ AD\ Leo} \times (\frac{T_\mathrm{eff\ *}}{T_\mathrm{eff\ AD\ Leo}})^4 \\ 
\times \frac{\textrm{log} (L_{H_\alpha *} / L_{bol *}) }{\textrm{log} (L_{H_\alpha AD Leo} / L_{bol AD Leo})} \nonumber
\end{eqnarray}
Our stellar temperature grid (Fig. \ref{Mstellarspectra}) covers the full M dwarf spectral range with the stellar parameters for active and inactive models given in Table \ref{tableM0toM9}. For the active star models, we scale the UV flux in the wavelength range 1000 - 3000\AA\ from AD Leo's observed UV spectrum to the other spectral types according to Eq. \ref{finalequation}. For each active model star on our grid we concatenated a solar metallicity, unreddened synthetic PHOENIX spectrum, which only considers photospheric emission \citep{allard2014}, from 3000 - 45,450\AA\ to the scaled observations from the International Ultraviolet Explorer (IUE) archive for AD Leo from 1000 - 3000\AA\ combined with the reconstructed Ly-$\alpha$ from \citet{linsky2013} for the region 1210 - 1222\AA.\footnote{http://archive.stsci.edu/iue/} Note that young active stars are highly variable, especially in the UV. AD Leo is the most active and well characterized M dwarf. Our active models therefore represent the extreme high end of activity for each stellar type from an active flare star. For comparison, the inactive semi-empirical models include no chromospheric emission. The continuum originating from the photosphere is taken from a PHOENIX model with the same stellar parameters as the corresponding active star from 1000 - 45,450\AA. Thus, these models provide only a lower limit to the stellar UV flux. More observations are needed to determine the true lower limit of inactive M dwarfs in the UV.

\renewcommand{\arraystretch}{0.6}
\begin{table}[h!]
\begin{center}
\caption{Stellar properties for active and inactive models\label{tableM0toM9}.}
\begin{tabular}{crrrrrrrrrrr}
\tableline\tableline
Star & $\teff$ (K) & Mass ($M_{\odot}$) & Radius ($R_{\odot}$)  \\
\tableline
M0 & 3800 & 0.60 & 0.62 \\
M1&	3600	 & 0.49	& 0.49\\
M2&3400&	0.44&	0.44\\
M3	&3200&	0.36	&0.39\\
M4&	3100	 &0.2&	0.36\\
M5&	2800	&0.14&	0.2\\
M6&	2600	&0.10&	0.15\\
M7&	2500	&0.09&	0.12\\
M8&	2400	&0.08&	0.11\\
M9&	2300	&0.075&	0.08\\
\tableline
\end{tabular}

\end{center}
\end{table}

The models represent the extreme limits of the UV radiation environment for an exoplanet using a flare star for the upper bound and a photosphere-only model for the lower bound in the UV.  Fig. \ref{photovschromo} shows the UV flux region between those two limits spans over 10 orders of magnitude in the FUV. We use the observations of six M dwarfs from the MUSCLES program to probe the region in between our active and inactive models. The MUSCLES stars have been traditionally classified as ``quiescent'' because they do not present H$_{\alpha}$ in emission. 

\begin{table*}[ht!]
\begin{center}
\caption{Stellar properties for MUSCLES model stellar spectra.\label{tableMuscles}}
\begin{tabular}{lccccccc}
\tableline\tableline
MUSCLES & $\teff$ (K) & Radius ($R_{\odot}$) & [Fe/H] & log $g$ & Age (Gyr) & $v$ sin $i$ \\
Stars & & & & & & (km s$^{-1}$) \\
\tableline
GJ 832 &	3620	&0.48	&-0.12	&4.70	&- &	$3^*$ \\
GJ 667C&	3350	&0.348	&-0.55	&5.00	& $>2$ &	$3^*$  \\
GJ 1214 &	3250	 & 0.211	& +0.05	& 4.99	&6$\pm$3 &	$<$ 1 \\
GJ 436&	3416	&0.455	&+0.04	&4.83	&6.5-9.9	& $<$ 1 \\
GJ 581&	3498	&0.299	&-0.10	&4.96	&7-11	& 2.1 \\
GJ 876&	3129	&0.3761	&+0.19	&4.89	&0.1-5	&1.38 \\
\tableline
\multicolumn{7}{l}{%
  \begin{minipage}{11cm}%
    \tiny Note: for GJ 832 and GJ 667C there is no $v$ sin $i$ measurement. We assumed a $v$ sin $i$ = 3.0 corresponding to the peak of the distribution \citep{jenkins2009} when generating a PHOENIX model. \\ See text for references.%
  \end{minipage}%
}\\
\end{tabular}
\end{center}
\end{table*}

The six M dwarfs were observed by two UV spectrographs (COS and STIS) on HST \citep{france2013}. We joined these HST measurements\footnote{http://cos.colorado.edu/~kevinf/muscles.html} with the star's corresponding PHOENIX models at 2800\AA\ after adjusting the HST flux levels to the level a planet would receive at the 1AU equivalent in the Habitable Zone. The input parameters for the PHOENIX models are the observed star's $\teff$, [Fe/H], log $g$, and the rotation velocity, $v$ sin $i$, as summarized in Table \ref{tableMuscles} and described for each star based on the most recent observations in the Appendix. When observations of $v$ sin $i$ were not available we used instead the peak of the observed distribution for 56 M dwarfs \citep{jenkins2009}. 

The MUSCLES database interpolates the region between 1760 - 2100 \AA\ because the noise in the observations was large in that region since those wavelengths were covered by the lowest sensitivity part of the STIS G230L bandpass.  In order to not bias the results too higher or too low, the MUSCLES team used the average flux in the 2100 - 2200 \AA\ region to interpolate the 1760 - 2100 \AA\ region. Follow up observations to the MUSCLES program with a new HST Treasury program aim to probe the 1760 - 2100 \AA\  region of the spectrum with higher sensitivity. We used the MUSCLES database and interpolation as given. Because this region of the UV is important for O$_3$ production, uncertainty of this interpolation could influence our atmospheric results.

Input active stellar spectra and MUSCLES spectra are shown in Fig. \ref{Mstellarspectra} (inactive models are not shown). 

\subsection{Planetary Atmosphere Model}

We use EXO-P \citep{kaltenegger2010a}, a coupled 1D radiative-convective atmosphere code developed for rocky exoplanets. The code incorporates a 1D climate \citep{kasting1986, pavlov2000, haqq2008}, 1D photochemistry \citep{pavlov2002, segura2005, segura2007}, and 1D radiative transfer model \citep{traub1976, kaltenegger2009} to calculate the model spectrum of an Earth-like exoplanet orbiting host M dwarfs in the Habitable Zone.

EXO-P is a model that simulates both the effects of stellar radiation on a planetary environment and the planet's outgoing spectrum. We model an altitude range that extends upwards to 60km with 100 height layers. We use a geometrical model in which the average 1D global atmospheric model profile is generated using a plane-parallel atmosphere, treating the planet as a Lambertian sphere, and setting the stellar zenith angle to 60 degrees to represent the average incoming stellar flux on the dayside of the planet \citep[see also][]{schindler2000}. The temperature in each layer is calculated from the difference between the incoming and outgoing flux and the heat capacity of the atmosphere in each layer. If the lapse rate of a given layer is larger than the adiabatic lapse rate, it is adjusted to the adiabatic rate until the atmosphere reaches equilibrium. We use a two-stream approximation \citep[see][]{toon1989}, which includes multiple scattering by atmospheric gases, in the visible/near IR to calculate the shortwave fluxes. Four-term, correlated-k coefficients parameterize the absorption by O$_3$, H$_2$O, O$_2$, and CH$_4$ \citep{pavlov2000}. In the thermal IR region, a rapid radiative transfer model (RRTM) calculates the longwave fluxes. Clouds are not explicitly calculated. The effects of clouds on the temperature vs. pressure profile are included by adjusting the surface albedo of the Earth-Sun system to have a surface temperature of 288K \citep[see][]{kasting1984, pavlov2000, segura2003, segura2005}. The photochemistry code, originally developed by \citet{kasting1985} solves for 55 chemical species linked by 220 reactions using a reverse-Euler method \cite[see][and references therein]{segura2010}. The photochemical model is stationary, and convergence is achieved when the following criteria are fulfilled: the production and loss rates of chemical species are balanced which results in a steady state for the chemical concentrations, and the initial boundary conditions, such a surface mixing ratios or surface fluxes, are met.

The radiative transfer model used to compute planetary spectra is based on a model originally developed for trace gas retrieval in Earth's atmospheric spectra \citep{traub1976} and further developed for exoplanet transmission and emergent spectra \citep{kaltenegger2007, kaltenegger2009, kaltenegger2010, kaltenegger2010a, kaltenegger2013}. In this paper, we model Earth's reflected and thermal emission spectra using 21 of the most spectroscopically significant molecules (H$_2$O, O$_3$, O$_2$, CH$_4$, CO$_2$, OH, CH$_3$Cl, NO$_2$, N$_2$O, HNO$_3$, CO, H$_2$S, SO$_2$, H$_2$O$_2$, NO, ClO, HOCl, HO$_2$, H$_2$CO, N$_2$O$_5$, and HCl). We use a Lambert sphere as an approximation for the disk integrated planet in our model. The surface of our model planet corresponds to Earth's current surface of 70\% ocean, 2\% coast, and 28\% land. The land surface consists of 30\% grass, 30\% trees, 9\% granite, 9\% basalt, 15\% snow, and 7\% sand. Surface reflectivities are taken from the USGS Digital Spectral Library\footnote{http://speclab.cr.usgs.gov/spectral-lib.html}  and the ASTER Spectral Library\footnote{http://speclib.jpl.nasa.gov}  \citep[following][]{kaltenegger2007}.

Clouds have a strong impact on the detectability of atmospheric species. For the spectra shown in Figs. \ref{VISspectraMstars}-\ref{methane}, we assume a 60\% global cloud cover with cloud layers analogous to Earth \citep[40\% water clouds at 1km, 40\% water clouds at 6km and 20\% ice clouds at 12km following][]{kaltenegger2007}. In the VIS to NIR, clouds increase the reflectivity of an Earth-like planet substantially and therefore increase the equivalent widths of all observable features, even though clouds block access to some of the lower atmosphere. In the IR, clouds slightly decrease the overall emitted flux of an Earth-like planet because they radiate at lower temperatures and therefore decrease the equivalent widths of all observable absorption features, even though they can increase the relative depth of a spectral feature due to lowering the continuum temperature of the planet. For a comparison of scenarios with Earth-analogue clouds to those of clear sky spectra see \citet{rugheimer2013}.

Using 34 layers, we calculate the spectrum at high spectral resolution with several points per line width. The line shapes and widths are computed using Doppler and pressure broadening on a line-by-line basis for each layer in the model atmosphere. The overall high-resolution spectrum is calculated with 0.1 cm$^{-1}$ wavenumber steps. The figures are shown smoothed to a resolving power of 150 in the IR and 800 in the VIS using a triangular smoothing kernel. The spectra may be binned further for comparison with proposed future spectroscopy missions designs to characterize Earth-like planets.  We previously validated EXO-P from the VIS to the IR using data from ground and space \citep{kaltenegger2007, kaltenegger2009, rugheimer2013}. 

\subsection{Simulation Set-Up}

To examine the effects of the varying UV flux of M dwarfs on an Earth-like atmosphere and its observable spectral features, we use the temperature grid of stellar models ranging from M9 to M0 ($\teff$ = 3800K to 2300K) for active and inactive models along with the six M dwarfs with well characterized UV fluxes from HST (see \S2.2). We simulated an Earth-like planet with the same mass as the Earth at the 1AU equivalent orbital distance, defined by the wavelength integrated stellar flux received on top of the planet's atmosphere being equivalent to 1AU in our solar system calculated by $1\ AU_{eq} = AU\times \sqrt{(R/R_{\odot})^2  \times (\teff / T_{\mathrm{eff}\ \odot})^4}$. 

The biogenic (produced by living organisms) fluxes were held fixed in the models in accordance with the fluxes that reproduce the modern mixing ratios in the Earth-Sun case, except for the cooler M dwarfs where CH$_4$ and N$_2$O were given a fixed mixing ratio of $1.0\times 10^{-3}$ and $1.5\times 10^{-2}$, respectively \citep[following][]{segura2003, segura2005}. The surface fluxes for the long-lived gases H$_2$, CH$_4$, N$_2$O, CO and CH$_3$Cl were calculated such that the Earth around the Sun yields a $\tsurf$ = 288K for surface mixing ratios: cH$_2$ = $5.5 \times 10^{-7}$, cCH$_4$ = $1.6 \times 10^{-6}$, cCO$_2$ = $3.5 \times 10^{-4}$, cN$_2$O = $ 3.0 \times 10^{-7}$, cCO = $9.0 \times 10^{-8}$, and cCH$_3$Cl = $5.0 \times 10^{-10}$ \citep[see][]{rugheimer2013}. The corresponding input surface fluxes to the atmosphere are $-1.9 \times 10^{12}$ g H$_2$ yr$^{-1}$, $5.3 \times 10^{14}$ g CH$_4$ yr$^{-1}$, $7.9 \times 10^{12}$ g N$_2$O yr$^{-1}$, $1.8 \times 10^{15}$ g CO yr$^{-1}$, and $4.3 \times 10^{12}$ g CH$_3$Cl yr$^{-1}$. The N$_2$ mixing ratio is set to be a ``fill gas'' such that the total surface pressure is 1 bar. These boundary conditions were used for M0-M5 active grid stars, M0-M3 inactive grid stars and all MUSCLES stars. 

For M6-M9 active grid stars and for M4-M9 inactive grid stars, the boundary condition for CH$_4$ is changed to a fixed mixing ratio once the UV environment of the host star drops below a certain level because, CO, H$_2$, and CH$_4$ do not converge according to the criteria described in \S2.2 assuming a modern Earth biological flux \citep[see also][]{segura2005}. In reality, the thermodynamical cost for microbes producing CH$_4$ would become unprofitable as temperatures would initially rise due the increased greenhouse effect, causing the microbes to operate less efficiently. This type of biological feedback is not included in the models. CH$_4$ is produced biotically by methanogens and other organisms and abiotically through hydrothermal vent systems. In the modern atmosphere there is a significant anthropogenic source of CH$_4$ from natural gas, livestock, and rice paddies. Upper limits of abiotic fluxes of methane can be estimated for terrestrial planets following \citet{guzman2014}. An estimate of Earth biotic methane is about 30x the abiotic flux \citep[see discussion in][]{segura2005}, and thus a range of methane fluxes may be maintained on a terrestrial planet via methanogenesis in addition to abiotic methane production.

Therefore, we set the mixing ratio of CH$_4$ to $1\times 10^{-3}$ corresponding to a value of the last stable run with an Earth-like biological CH$_4$ flux for an active M5 star for planets around host stars that show this ``run-away'' behavior. Given these boundary conditions, the methane flux necessary to sustain a $1\times 10^{-3}$ mixing ratio of methane is $5.65\times 10^{14}$ g yr$^{-1}$ (equal to our present Earth methane flux) for the planet around the hottest M dwarf ($\teff$ = 3800 K) either active or inactive, $1.16\times 10^{14}$ g yr$^{-1}$ (20.5\% of the Earth value) for the planet around the coolest ($\teff$ = 2300 K) active star, and $5.87\times 10^{12}$ g yr$^{-1}$ (1\% of the Earth value) for the coolest inactive M dwarf of our sample.

For the less active stars, we set the deposition velocity (rate at which they are deposited to the surface) for H$_2$ and CO to $2.4\times 10^{-4}$ cm s$^{-1}$ and $1.2\times 10^{-4}$ cm s$^{-1}$, respectively \citep[see also][]{segura2005, kharecha2005}. 
\citet{rauer2011} used a deposition velocity for  H$_2$ of $7.7\times 10^{-4}$ cm s$^{-1}$. This number was obtained after calibrating the model to reproduce the  H$_2$ atmospheric abundance measured for present Earth. In our case, the deposition values for H$_2$ and CO correspond to maximum air-sea transfer rates estimated using the ``piston velocity'' approach \citep{broecker1982, kharecha2005}. Since a larger H$_2$ deposition velocity implies consumption by bacteria, we use the abiological value in \citet{kharecha2005}.

We followed a similar procedure for constraining the N$_2$O concentrations for M6-M9 inactive stars. We set the mixing ratio of N$_2$O to $1.5\times 10^{-2}$, corresponding to values of the last stable run with Earth-like biological fluxes, to prevent unphysical buildup of N$_2$O. N$_2$O is emitted primarily by (de)nitrifying bacteria with anthropogenic sources from fertilizers in agriculture, biomass burning, industry, and livestock and is a relatively minor constituent of the modern atmosphere at around 320 parts per billion (ppb), compared to the pre-industrial concentration of 270 ppb \citep{forster2007}. 

We used modern Earth fluxes for all of the MUSCLES stars without any ``run-away'' buildup of reduced gases. The CH$_4$ and N$_2$O fluxes are given in Table \ref{tableCH4N2O}. 

All of our simulations used a fixed mixing ratio of 355ppm for CO$_2$,  21\% O$_2$, and a fixed upper boundary of $10^{-4}$ bar ($\sim$ 60km). In the 1D climate model, a surface albedo of 0.2 is fixed for all simulations, corresponding to the surface albedo that reproduces Earth's average temperature of 288K for the Earth/Sun case. The planetary Bond albedo (surface + atmosphere) is calculated by the 1D climate code.cm$^{-2}$

\renewcommand{\arraystretch}{0.6}
\begin{table*}[ht!]
\begin{center}
\caption{CH$_4$ and N$_2$O fluxes and mixing ratios and O$_3$ column depths.\label{tableCH4N2O}}
\begin{tabular}{lccccccc}
\tableline\tableline
  & Surface            & Flux (g/yr) & \% Earth       & Surface            & Flux (g/yr) & \% Earth       & O$_3$ Column  \\
  &  Mixing             & CH$_4$    & CH$_4$ flux & Mixing              & N$_2$O    & N$_2$O flux & Depth  \\
  &  Ratio CH$_4$ &                  &                      & Ratio N$_2$O &                  &                & (cm$^{-2}$) \\
\tableline

Active  \\	
\hline						
M0 A	 & 330ppm &	$5.65\times 10^{14}$ & 	100\% & 	0.70ppm & 	$7.99\times 10^{12}$ & 	100\% & 	$4.1\times 10^{18}$ \\	
M1 A & 	370ppm&	$5.65\times 10^{14}$ & 	100\% & 	0.70ppm & 	$7.99\times 10^{12}$ & 	100\% & 	$4.0\times 10^{18}$ \\	
M2 A & 	450ppm&	$5.65\times 10^{14}$ & 	100\% & 	0.76ppm & 	$7.99\times 10^{12}$ & 	100\% & 	$3.8\times 10^{18}$ \\	
M3 A & 	570ppm&	$5.65\times 10^{14}$ & 	100\% & 	0.82ppm & 	$7.99\times 10^{12}$ & 	100\% & 	$3.6\times 10^{18}$ \\	
M4 A & 	910ppm&	$5.65\times 10^{14}$ & 	100\% & 	0.93ppm & 	$7.99\times 10^{12}$ & 	100\% & 	$3.5\times 10^{18}$ \\	
M5 A	 & 1000 ppm& $5.65\times 10^{14}$	 & 100\% & 	0.98 ppm & 	$7.99\times 10^{12}$ & 	100\% & 	$2.6\times 10^{18}$ \\	
M6 A	 & 1000ppm	&$2.90\times 10^{14}$ & 	51.8\% & 	1.7ppm & 	$7.99\times 10^{12}$ & 	100\% & 	$3.1\times 10^{18}$ \\	
M7 A	 & 1000ppm	&$1.28\times 10^{14}$ & 	22.7\% & 	3.0ppm & 	$7.99\times 10^{12}$ & 	100\% & 	$2.4\times 10^{18}$ \\	
M8 A	 & 1000ppm	&$1.36\times 10^{14}$ & 	24.0\% & 	3.1ppm & 	$7.99\times 10^{12}$ & 	100\% & 	$2.5\times 10^{18}$ \\	
M9 A	 & 1000ppm	&$1.20\times 10^{14}$ & 	21.2\% & 	3.5ppm & 	$7.99\times 10^{12}$ & 	100\% & 	$2.3\times 10^{18}$ \\	
\hline
Inactive							 \\	
\hline
M0 I	 & 490 ppm&	$5.65\times 10^{14}$ & 	100\% & 	34 ppm & 	$7.99\times 10^{12}$ & 	100\% & 	$1.7\times 10^{18}$ \\	
M1 I	 & 580 ppm&	$5.65\times 10^{14}$ & 	100\% & 	64 ppm & 	$7.99\times 10^{12}$ & 	100\% & 	$1.5\times 10^{18}$ \\	
M2 I	 & 650 ppm&	$5.65\times 10^{14}$ & 	100\% & 	120 ppm & 	$7.99\times 10^{12}$ & 	100\% & 	$1.4\times 10^{18}$ \\	
M3 I	 & 1000 ppm&$5.65\times 10^{14}$ & 	100\% & 	360 ppm & 	$7.99\times 10^{12}$ & 	100\% & 	$1.3\times 10^{18}$ \\	
M4 I	 & 1000ppm	&$4.15\times 10^{14}$	 & 73.5\% & 	670ppm	& $7.99\times 10^{12}$ & 	100\% & 	$1.0\times 10^{18}$ \\	
M5 I	 & 1000ppm	&$1.23 \times 10^{14}$ & 	21.8\% & 	15000ppm  & 	$7.99\times 10^{12}$ & 	100\% & 	$4.5\times 10^{17}$ \\	
M6 I	 & 1000ppm	&$2.42 \times 10^{13}$ & 	4.3\% & 	15000ppm & 	$4.04\times 10^{11}$	 &  5.1\% & 	$1.4\times 10^{17}$ \\	
M7 I	 & 1000ppm	&$1.67 \times 10^{13}$ & 	3.0\% & 	15000ppm & 	$4.96\times 10^{10}$ & 	0.6\% & 	$6.1\times 10^{16}$ \\	
M8 I	 & 1000ppm	&$1.34\times 10^{13}$ & 	2.4\% & 	15000ppm & 	$5.72\times 10^{9}$	 & 0.07\%	 & $2.6\times 10^{16}$ \\	
M9 I	 & 1000ppm	&$5.87 \times 10^{12}$ & 	1.0\% & 	15000ppm & 	$3.17\times 10^{9}$	 & 0.04\%	 & $1.1\times 10^{16}$ \\	
\hline
MUSCLES							 \\	
\hline
GJ 832	 & 550ppm	&$5.65\times 10^{14}$ & 	100\% & 	15ppm & 	$7.99\times 10^{12}$ & 	100\% & 	$1.6\times 10^{18}$ \\	
GJ 667C	 & 330ppm	&$5.65\times 10^{14}$ & 	100\% & 	0.7ppm & 	$7.99\times 10^{12}$ & 	100\% & 	$7.0\times 10^{18}$ \\	
GJ 1214	 & 1600ppm	&$5.65\times 10^{14}$ & 	100\% & 	1.1ppm & 	$7.99\times 10^{12}$ & 	100\% & 	$2.3\times 10^{18}$ \\	
GJ 436	 & 1600ppm	&$5.65\times 10^{14}$ & 	100\% & 	4.3ppm & 	$7.99\times 10^{12}$ & 	100\% & 	$1.1\times 10^{18}$ \\	
GJ 581	 & 1400ppm	&$5.65\times 10^{14}$ & 	100\% & 	2.9ppm & 	$7.99\times 10^{12}$ & 	100\% & 	$1.2\times 10^{18}$ \\	
GJ 876	 & 3400ppm	&$5.65\times 10^{14}$ & 	100\% & 	6.9ppm & 	$7.99\times 10^{12}$ & 	100\% & 	$8.8\times 10^{17}$ \\	
\tableline
\end{tabular}
\end{center}
\end{table*}
\section{ATMOSPHERIC MODEL RESULTS}

The amount of UV radiation emitted from the host star influences the abundances of major chemical atmospheric constituents including H$_2$O, CH$_4$, and O$_3$ and, as a result, modified the temperature-pressure profile of a planet. The UV fluxes incident at the top of the atmosphere of an Earth-like planet in the HZ are given in Table \ref{tableUVFluxes} and shown in Fig. \ref{Mstellarspectra}. An M0 active star model has 6.7 times more total UV flux than an M0 inactive star model. An M9 active star model has 4600 times more total UV flux than an M9 inactive star model. The greatest differences are in the far UV (FUV 1000 - 2000 \AA) where an active star model has $6.8 \times 10^4$ and $1.3 \times 10^{11}$ more FUV flux than an inactive M0 and M9 model, respectively. The MUSCLES stars' UV environments fall between the active and inactive star models, consistent with their classification as being weakly active.

\begin{table*}[ht!]
\begin{center}
\caption{Integrated UV fluxes at the top of the atmosphere (TOA).\label{tableUVFluxes}}
\begin{tabular}{lccccc}
\tableline\tableline
Stellar & Ly-$\alpha$ TOA  & FUV minus Ly-$\alpha$  & FUV TOA         & NUV TOA        & Full UV TOA\\
Type  &  1210-1222\AA      & TOA 1222-2000\AA\       & 1000-2000\AA  & 2000-3200\AA  & 1000-3200\AA \\
         & ergs cm$^{-2}$ s$^{-1}$      &    ergs cm$^{-2}$ s$^{-1}$           &ergs cm$^{-2}$ s$^{-1}$  &ergs cm$^{-2}$ s$^{-1}$ &ergs cm$^{-2}$ s$^{-1}$ \\
\hline
Active	\\
\hline				
M0 A &  $6.9 \times 10^2$ & $2.4 \times 10^2$ & $9.4 \times 10^2$ &	$1.4 \times 10^3$ & 	$2.3 \times 10^3$ \\
M1 A & 	$5.6 \times 10^2$ & $2.0 \times 10^2$ & $7.6 \times 10^2$ &	$1.1 \times 10^3$ &	$1.9 \times 10^3$ \\
M2 A & 	$4.4 \times 10^2$ & $1.6 \times 10^2$ & $6.1 \times 10^2$ &	$8.5 \times 10^2$ &	$1.5 \times 10^3$ \\
M3 A &  $3.5 \times 10^2$ & $1.2 \times 10^2$ & $4.7 \times 10^2$ &	$6.5 \times 10^2$ &	$1.1 \times 10^3$ \\
M4 A & 	$3.1 \times 10^2$ & $1.1 \times 10^2$ & $4.2 \times 10^2$ &	$5.6 \times 10^2$ &	$9.8 \times 10^2$ \\
M5 A & 	$1.8 \times 10^2$ & $6.3 \times 10^1$ & $2.5 \times 10^2$ &	$3.3 \times 10^2$ &	$5.8 \times 10^2$ \\
M6 A & 	$6.4 \times 10^1$ & $2.3 \times 10^1$ & $8.8 \times 10^1$ &	$1.2 \times 10^2$ &	$2.1 \times 10^2$ \\
M7 A & 	$2.6 \times 10^1$ & $9.2 \times 10^0$ & $3.6 \times 10^1$ &	$4.8 \times 10^2$ &	$8.4 \times 10^2$ \\
M8 A &  $2.6 \times 10^1$ & $9.0 \times 10^0$ & $3.5 \times 10^1$ &	$4.7 \times 10^2$ &	$8.2 \times 10^2$ \\
M9 A &  $2.2 \times 10^1$ & $7.6 \times 10^0$ & $3.0 \times 10^1$ &	$3.9 \times 10^2$ &	$6.9 \times 10^2$ \\
\hline
Inactive \\
\hline					
M0 I	 & $2.7 \times 10^{-11}$ &	$9.4 \times 10^{-3}$ &	$9.5 \times 10^{-3}$ &	$3.0 \times 10^2$ &	$3.0 \times 10^2$ \\
M1 I	 & $5.8 \times 10^{-12}$ &	$2.9 \times 10^{-3}$ &	$2.9 \times 10^{-3}$ &	$2.6 \times 10^2$ &	$2.6 \times 10^2$ \\
M2 I	 & $1.5 \times 10^{-12}$ &	$9.1 \times 10^{-4}$ &	$9.1 \times 10^{-4}$ &	$1.7 \times 10^2$ &	$1.7 \times 10^2$ \\
M3 I	 & $3.5 \times 10^{-14}$ &	$6.9 \times 10^{-5}$ &	$6.9 \times 10^{-5}$ &	$8.9 \times 10^1$ & 	$8.9 \times 10^1$ \\
M4 I	 & $6.1 \times 10^{-15}$ & 	$2.2 \times 10^{-5}$ &	$2.2 \times 10^{-5}$ &	$5.9 \times 10^1$ &	$5.9 \times 10^1$ \\
M5 I	 & $2.9 \times 10^{-19}$ &	$1.5 \times 10^{-8}$ &	$1.5 \times 10^{-8}$ &	$1.3 \times 10^1$ &	$1.3 \times 10^1$ \\
M6 I	 & $4.2 \times 10^{-21}$ &	$2.6 \times 10^{-9}$ &	$2.6 \times 10^{-9}$ &	$3.0 \times 10^0$ &	$3.0 \times 10^0$ \\
M7 I	 & $1.1 \times 10^{-21}$ &	$2.6 \times 10^{-10}$ &	$2.6 \times 10^{-10}$ &	$1.3 \times 10^0$ &	$1.3 \times 10^0$ \\
M8 I	 & $8.3 \times 10^{-24}$ &	$1.7 \times 10^{-10}$ &	$1.7 \times 10^{-10}$ &	$5.0 \times 10^{-1}$ &	$5.0 \times 10^{-1}$ \\
M9 I	 & $1.8 \times 10^{-24}$ &	$1.6 \times 10^{-10}$ &	$1.6 \times 10^{-10}$ &	$1.3 \times 10^{-1}$ &	$1.3 \times 10^{-1}$ \\
\hline
MUSCLES	 \\
\hline				
GJ 832 &	$1.5 \times 10^2$ &	$3.0 \times 10^0$ &	$1.5 \times 10^2$ &	$3.5 \times 10^2$ &	$5.0 \times 10^2$ \\
GJ 667C	 &      $1.2 \times 10^2$ &	$2.4 \times 10^0$ &	$1.3 \times 10^2$ &	$2.3 \times 10^2$ &	$3.6 \times 10^2$ \\
GJ 1214 &	$1.4 \times 10^0$ &	$9.9 \times 10^1$ &	$1.1 \times 10^2$ &	$1.8 \times 10^2$ & 	$2.8 \times 10^2$ \\
GJ 436 &	$6.1 \times 10^1$ &	$7.1 \times 10^0$ &	$6.8 \times 10^1$ &	$2.3 \times 10^2$ &	$2.9 \times 10^2$ \\
GJ 581 &	$4.2 \times 10^1$ &	$6.8 \times 10^0$ &	$5.0 \times 10^1$ &	$2.7 \times 10^2$ &	$3.2 \times 10^2$ \\
GJ 876 &	$3.4 \times 10^1$ &	$1.1 \times 10^1$ &	$4.9 \times 10^1$ &	$1.1 \times 10^2$ &	$1.6 \times 10^2$ \\
\tableline
\tableline
\end{tabular}
\end{center}
\end{table*}

\begin{figure*}[ht!]
\centering
\includegraphics[scale=0.65,angle=-90]{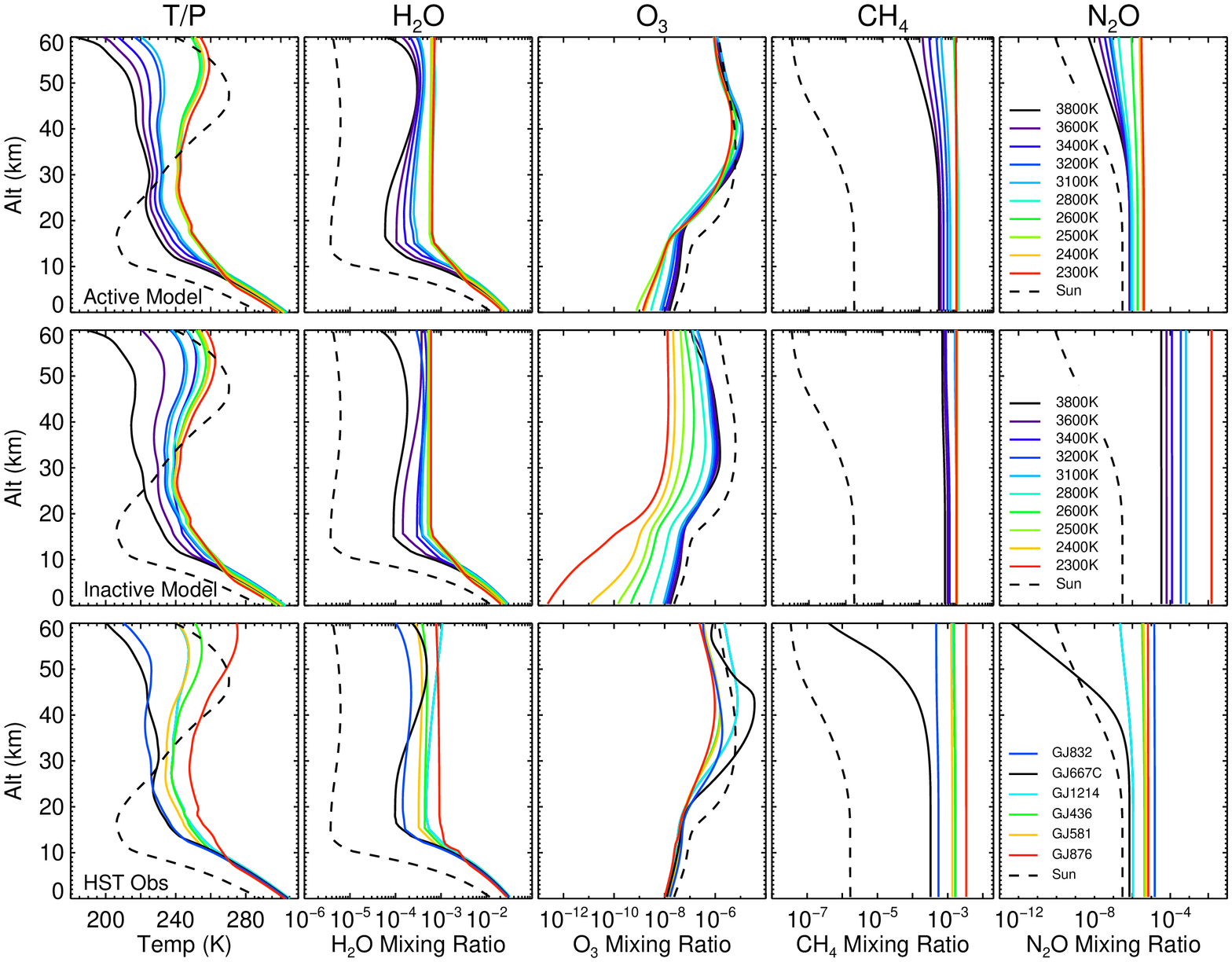}
\caption{Planetary temperature vs. altitude profiles and mixing ratio profiles profiles for H$_2$O, O$_3$, CH$_4$ and N$_2$O (left to right) for an Earth-like planet orbiting the grid of active stellar models (top), inactive stellar models (middle), and MUSCLES stars (bottom). Earth-Sun profiles are shown for comparison (dashed black lines).\label{tpchemMstars}}
\end{figure*}

The temperature vs altitude profile and the H$_2$O, O$_3$, CH$_4$ and N$_2$O mixing ratio profiles for all of the simulations are shown in Fig. \ref{tpchemMstars}, with each row corresponding to active, inactive, and MUSCLES star models, respectively. CH$_3$Cl profiles are not shown here, but follow the same trends as CH$_4$. Since both O$_2$ and CO$_2$ are well mixed in the atmosphere, their vertical mixing ratio profiles of 0.21 and 355ppm, respectively, are not shown. 

In the first column of Fig. \ref{tpchemMstars}, we show the changes in the temperature/altitude profile for Earth-like atmosphere models around M dwarfs for active (top row), inactive (middle row) and the six observed MUSCLES stars (bottom row). All temperature inversions are weaker than for the modern Earth because M dwarfs emit low UV flux in the 2000 - 3000 Å wavelength region, thereby producing near isothermal stratospheres \citep[see also][]{segura2005}. Fig. \ref{tpchemMstars} shows that temperature inversions are weaker for the higher UV environment stars. This is counter-intuitive since for the modern Earth, O$_3$ absorption of UV radiation causes stratospheric heating and an inversion. However in these lower UV environments, additional heating is provided by stratospheric CH$_4$ and H$_2$O. 

H$_2$O concentrations are lower in the stratosphere for planets around M dwarf models with higher UV fluxes, although higher O$_3$ concentrations will act as a shield, partially offsetting the effect of higher UV photon fluxes (Fig. \ref{tpchemMstars}, 2nd column). H$_2$O can also be formed in the stratosphere from CH$_4$ and OH (CH$_4$ + OH $\rightarrow$ CH$_3$ + H$_2$O) and by increased upwards vertical transport in the nearly isothermal stratospheres of Earth-like planets orbiting M dwarfs \citep[see also][]{segura2005}. O$_3$ shields H$_2$O in the troposphere from UV environments. While photochemically inert in the troposphere, H$_2$O can be removed by photolysis at wavelengths shortward of 2000\AA\ in the stratosphere (see columns 2 \& 3 in Table \ref{tableUVFluxes}) or by reactions with excited oxygen, O($^1$D), to produce OH radicals. 

In an atmosphere containing O$_2$, O$_3$ concentrations are determined primarily by the absorption of UV light shortward of 2400\AA\ in the stratosphere, and we see a corresponding decrease in O$_3$ concentrations correlated with decreased FUV radiation for cooler M dwarfs as well as inactive versus active stars (Fig. \ref{tpchemMstars}, 3rd column). 

CH$_4$ mixing ratio profiles are shown in column 4 of Fig. \ref{tpchemMstars}. As mentioned previously, we set the mixing ratio of methane to be 1000 ppm for the later M dwarf spectral types as described in \S2.3 and Table \ref{tableCH4N2O}. For Earth-like CH$_4$ fluxes, CH$_4$ concentrations decrease with higher UV environments due to reactions with OH to form H$_2$O and by photolysis by $\lambda <$ 1500 \AA\ (Fig. \ref{tpchemMstars}, 4th column). 

N$_2$O mixing ratios are larger for cooler M dwarfs because cooler stars have smaller UV fluxes at $\lambda <$ 2200 \AA and thus lower photolysis rates. Around later M dwarfs, we see mixing ratios of N$_2$O that increase unrealistically for the inactive stellar models M6-M9, similar to CH$_4$ (Fig. \ref{tpchemMstars}, 5th column). Since such a ``run-away'' effect should not be sustainable by biology, we cap the N$_2$O to a fixed mixing ratio corresponding to the last stable run for the inactive model M5. N$_2$O is also an indirect sink for stratospheric O$_3$, since about 1\% is converted to NO. Therefore, increasing N$_2$O decreases O$_3$ abundance. 

OH concentrations decrease with decreasing UV levels (in cooler and inactive M dwarfs) and OH is a primary sink for many species in the atmosphere including, but not limited to, many biologically interesting species including CH$_4$, CH$_3$, HCl, H$_2$, H$_2$S and CH$_3$Cl. CH$_3$Cl concentrations increase for cooler and inactive M dwarfs due to decreased stellar UV flux. 

O$_2$ and CO$_2$ concentrations remain constant and well mixed for all stellar types.

\begin{figure}[h!]
\centering
\includegraphics[scale=0.37,angle=-90]{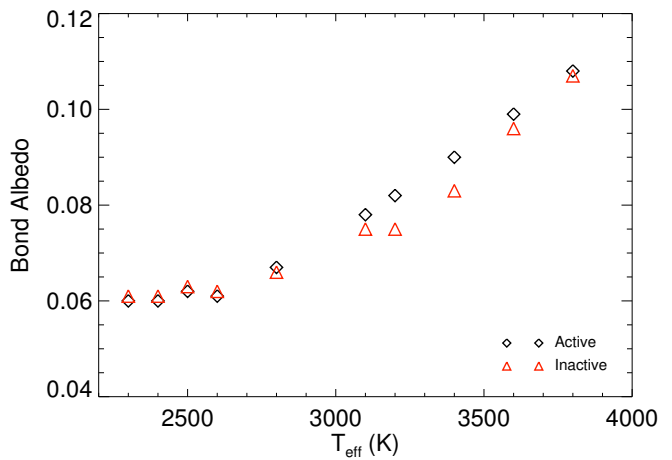}
\caption{Calculated Bond albedos for Earth-like planets orbiting the M0-M9 Active (black diamond) and Inactive (red triangle) stellar models. \label{albedoMstars}}
\end{figure}

The planetary surface temperatures range between 297 - 304K and Bond albedos range between 0.108 - 0.06 for M0 to M9 stars, respectively (see Fig. \ref{albedoMstars}). These albedos are lower than Earth's Bond albedo of 0.3 around the Sun because the stellar spectral energy distributions peak at longer wavelengths for cooler stars where Rayleigh scattering is less efficient, assuming the same total insolation.

\subsection{Effect of Ly-$\alpha$}

Ly-$\alpha$ is the brightest line in the UV spectra for cool stars and accounts for a significant portion of the overall UV flux. However, the intrinsic Ly-$\alpha$ flux must be reconstructed to account for interstellar absorption by neutral hydrogen \citep[see][]{wood2005, linsky2013}. For the observed MUSCLES stars, the Ly-$\alpha$ line flux is 13-33\% the total UV flux excluding Ly-$\alpha$. GJ 1214 has no observed Ly-$\alpha$ and only an upper limit which is 0.5\% of the total UV flux \citep{france2013}. In our active models, Ly-$\alpha$ ranges from 2 to 17\% of the total UV flux based on observations of AD Leo. In the inactive models (i.e. no chromosphere) Ly-$\alpha$ is negligible compared with total UV flux.

\begin{figure*}[ht!]
\centering
\includegraphics[scale=0.65,angle=-90]{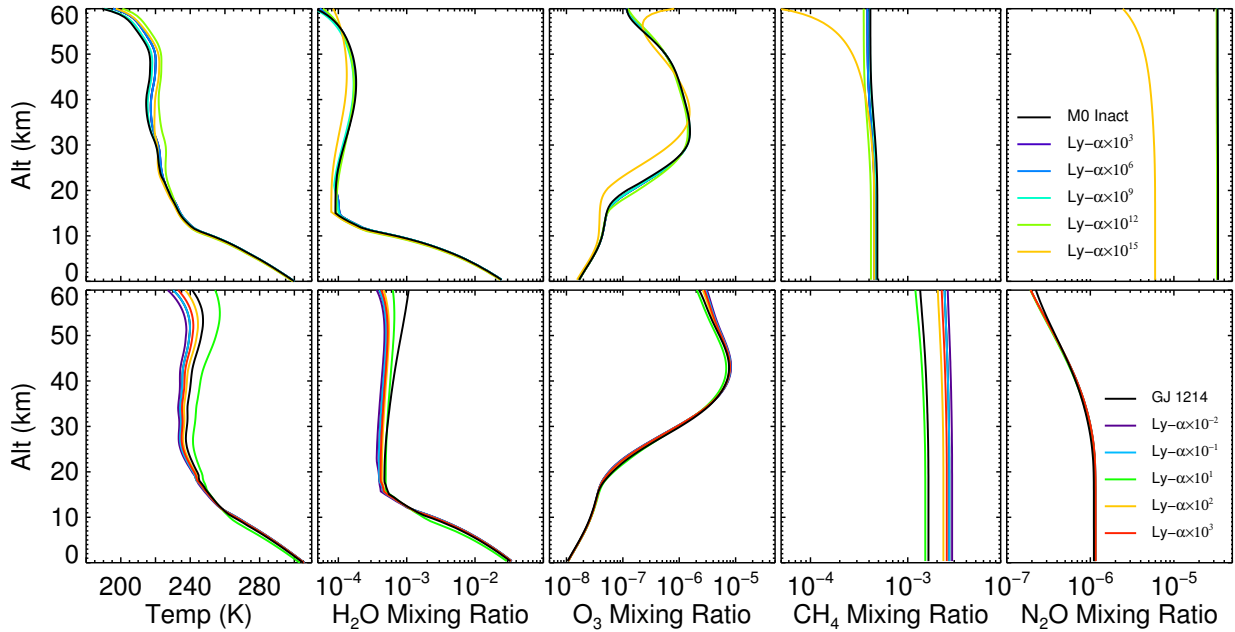}
\caption{Temperature, H$_2$O, O$_3$, CH$_4$, and N$_2$O mixing ratio profiles where the Ly-$\alpha$ has been increased by a factor of $10^3$, $10^6$, $10^9$, $10^{12}$ and $10^{15}$ above the inactive M0 model star corresponding to flux levels of $2.7\times 10^{-8}$, $2.7\times 10^{-5}$, $2.7\times 10^{-2}$, $2.7\times 10^{1}$, $2.7\times 10^{4}$ ergs cm$^{-2}$ s$^{-1}$, respectively (top) and for GJ 1214 where the Ly-$\alpha$ upper limit ($2.4\times 10^{-15}$ ergs cm$^{-2}$ s$^{-1}$) has been multiplied by a factor of $10^{-2}$, $10^{-1}$, $10{^1}$, $10^2$ and $10^3$ (bottom) \label{tpchemlya}}
\end{figure*}

We ran two sensitivity tests of the photochemistry of Earth-like planets to the amount of Ly-$\alpha$ flux from their host star. First we increased the Ly-$\alpha$ by a factor of $10^3$, $10^6$, $10^9$, $10^{12}$ and $10^{15}$ above M0 inactive model corresponding to flux levels of $2.7\times 10^{-8}$, $2.7\times 10^{-5}$, $2.7\times 10^{-2}$, $2.7\times 10^{1}$, $2.7\times 10^{4}$ ergs cm$^{-2}$ s$^{-1}$ respectively. The highest Ly-$\alpha$ value considered is 84x higher than that of the M0 active model and 180x higher than the highest observed Ly-$\alpha$ flux in the MUSCLES stellar sample. As seen in Fig. \ref{tpchemlya} (top), increasing the Ly-$\alpha$ flux has only a small effect on the photochemistry until the most extreme case considered where it primarily photolyzes N$_2$O and CH$_4$. We run our models to around 60km corresponding to a pressure of $1 \times 10^{-4}$ bar. The effect of Ly-$\alpha$ will be more pronounced at pressures lower than this as \citet{miguel2014} has shown for mini-neptune atmospheres.

GJ 1214 is the only MUSCLES star to not have a directly detected Ly-$\alpha$ flux. An upper limit was placed at $2.4\times 10^{-15}$ ergs cm$^{-2}$ s$^{-1}$. We artificially set the Ly-$\alpha$ flux to be $10^{-2}$, $10^{-1}$, $10^1$, $10^2$, and $10^3$ times this upper limit. We tested values both above and below the upper limit to see how sensitive an Earth-like planet atmosphere is to changes in Ly-$\alpha$ flux when the rest of the NUV radiation field is still much higher than the inactive model in the previous sensitivity test. As shown in Fig. \ref{tpchemlya} (bottom), the largest changes are seen in the CH$_4$ and to a smaller extent in stratospheric H$_2$O concentrations. O$_3$ and N$_2$O remain relatively constant through the 5 orders of magnitude change in Ly-$\alpha$ flux. 

Figures \ref{tpchemMstars} and \ref{tpchemlya} show it is important to characterize the entire UV spectrum including the near UV and the base level flux between emission lines, to understand future observations of extrasolar planet atmospheres. Ly-$\alpha$ is one of the most important lines to characterize.

\section{RESULTS: SPECTRA OF EARTH-LIKE PLANETS ORBITING M DWARFS}

Spectra of Earth-like planets orbiting M dwarfs with varying UV activity levels show measureable differences in spectral feature depths. In the VIS, the depth of absorption features is primarily sensitive to the abundance of the species, while in the IR, both the abundance and the temperature difference between the emitting/absorbing layer and the continuum influence the depth of features.
 
We assume full phase (secondary eclipse) for all spectra presented to show the maximum flux that can be observed. Figs \ref{VISspectraMstars}-\ref{contrastMstars} show an Earth-size planet to determine the specific flux and planet-to-star contrast ratio. A Super-Earth with up to twice Earth's radius will provide 4 times more flux and a better contrast ratio. No noise has been added to these model planetary spectra to provide inputs for a wide variety of instrument simulators for both secondary eclipse and direct detection missions.

\subsection{Earth-like Visible/Near-infrared Spectra (0.4$\mu$m - 4$\mu$m)}

\begin{figure*}[ht!]
\centering
\includegraphics[scale=0.65]{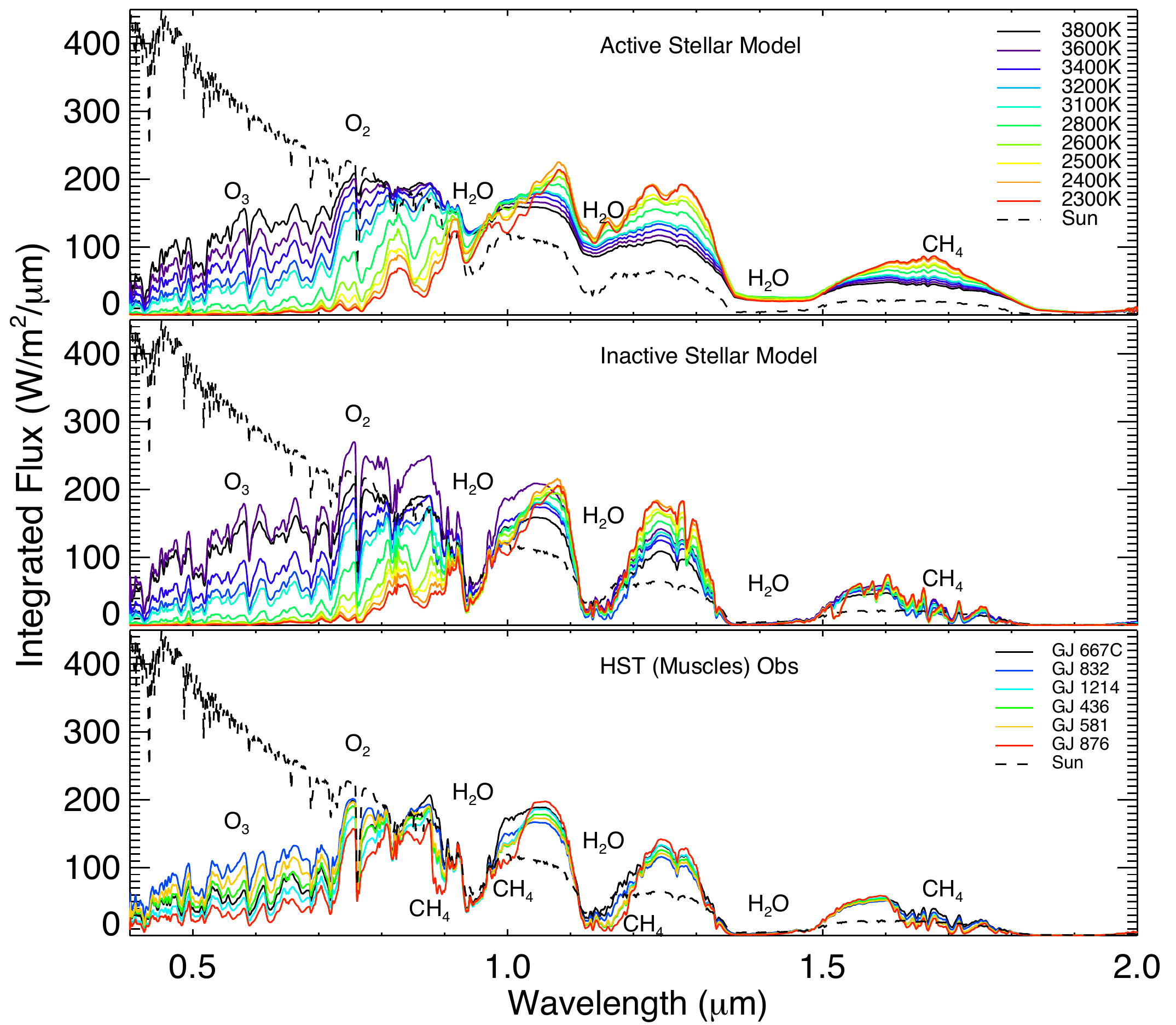}
\caption{Smoothed, disk-integrated VIS/NIR spectra at the TOA for an Earth-like planet for the grid of active stellar models (top), inactive stellar models (middle), and MUSCLES stars (bottom) assuming 60\% Earth-analogue cloud coverage model (region 2-4$\mu$m has low integrated flux levels and therefore is not shown here). The Earth-Sun spectrum is shown for comparison as a dashed black line. \label{VISspectraMstars}}
\end{figure*}

Fig. \ref{VISspectraMstars} shows the reflected visible and near-infrared spectra from 0.4 to 2$\mu$m of Earth-like planets around the M dwarf grid of active, inactive, and MUSCLES stars, using the SED of the host star. We assume Earth-analogue cloud cover. The high-resolution spectra calculated with 0.1 cm$^{-1}$ steps have been smoothed to a resolving power of 800 using a triangular smoothing kernel to show the individual features more clearly. The Earth-Sun spectrum is shown for comparison as a dashed black line. 

\begin{figure}[h!]
\centering
\includegraphics[scale=0.34]{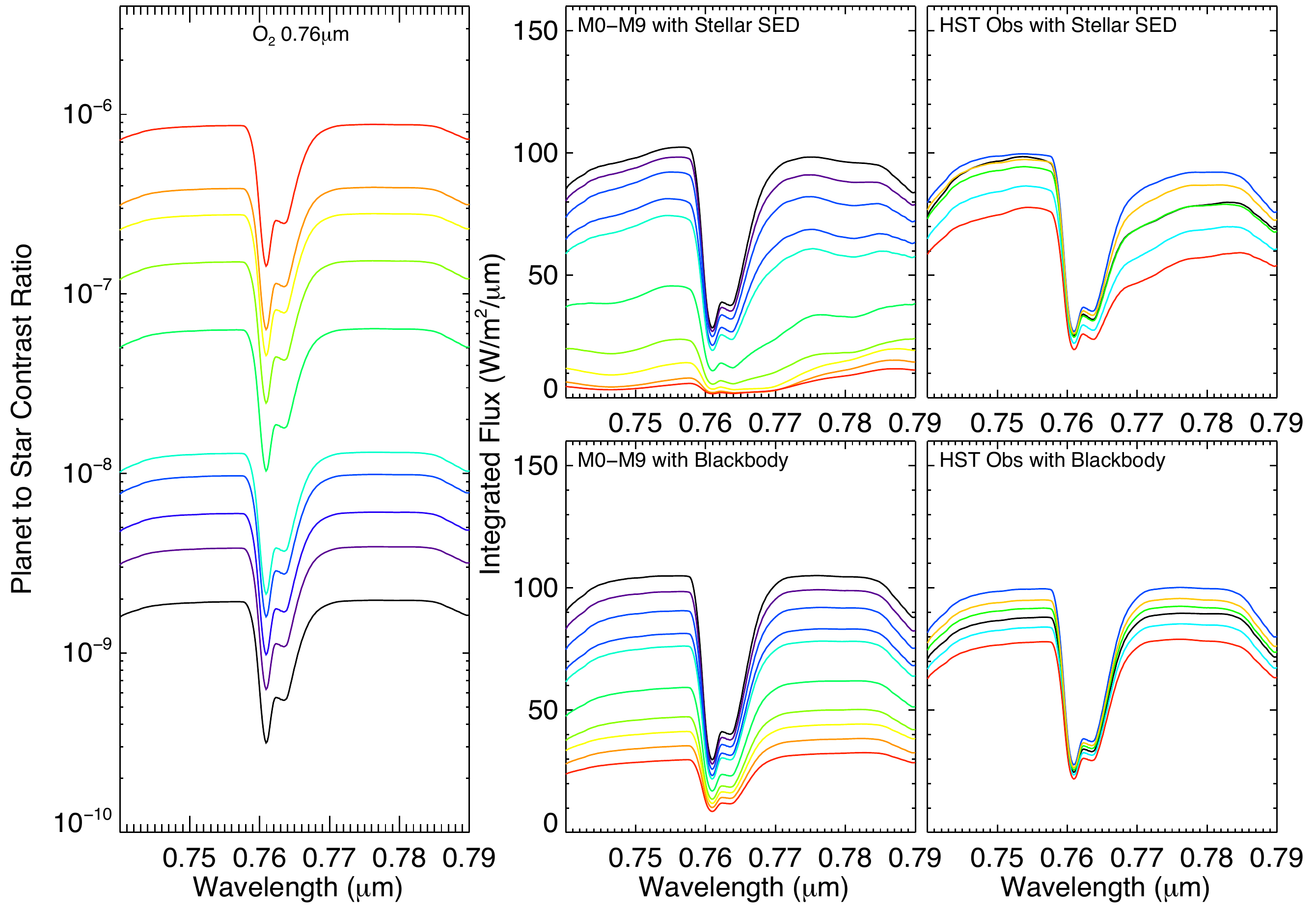}
\caption{O$_2$ feature at 0.76$\mu$m relative flux as planet-to-star contrast ratio (left), and the reflected emergent flux for a 60\% cloud cover model for M0-M9 model stars (top middle), and for MUSCLES stars (top right). Bottom row shows the same but with a blackbody used to calculate the absolute flux. The observable O$_2$ feature becomes increasingly difficult to detect for later stellar types due to the decreasing stellar flux in this wavelength region. Coloring is same as in Fig. \ref{VISspectraMstars}. \label{oxygenmstars}}
\end{figure}

Due to the increased stellar flux at shorter wavelengths for a G-type star, Rayleigh scattering is much more pronounced for FGK stars, which greatly increases the flux from 0.4 to 0.8$\mu$m for an Earth-like planet around a hotter star. Since M dwarfs have stronger NIR emission, and the 1-2$\mu$m flux is larger around Earth-like planets orbiting M dwarfs than for the Earth-Sun equivalent. The most notable features in the VIS/NIR spectra are O$_3$ at 0.6$\mu$m (the Chappuis band), O$_2$ at 0.76$\mu$m, H$_2$O at 0.95$\mu$m, and CH$_4$ at 0.6, 0.7, 0.8, 0.9, 1.0 and 1.7$\mu$m. Note that shallow spectral features like the visible O$_3$ feature would require a very high signal-to-noise ratio (SNR) to be detected.

Fig. \ref{oxygenmstars} shows details for one of the most notable feature in the VIS, the O$_2$ A band at 0.76$\mu$m in both the relative flux as planet-to-star contrast ratio (left), and the reflected emergent flux for a 60\% cloud cover model for M0-M9 model stars (top middle), and for the six MUSCLES stars (top right). Note that the detectability of the O$_2$ feature in reflected light is similar for active and inactive models since the stellar flux at 0.76$\mu$m is activity independent. The relative flux shows an equivalently deep feature for each case due to a constant mixing ratio of 21\%. However, the oxygen feature in absolute flux (Fig. 8, upper middle panel) becomes increasingly difficult to detect for the later stellar types. For the latest M stellar types modeled here, the detection of the O$_2$ feature requires a very high SNR to detect even if the planet has an active photosynthetic biosphere like the Earth (see Fig. \ref{VISspectraMstars}). If one assumes a blackbody radiation Planck function, rather than a realistic stellar model, when calculating the shape of the reflected light curve, the reduction in the feature's depth on moving to later stellar types (i.e. going from blue to red lines in Fig. \ref{oxygenmstars}, 2nd and 3rd columns) is less pronounced. The O$_2$ feature is pronounced for all MUSCLES stars since none of those stars have a low enough $\teff$ for the O$_2$ feature to be diminished by the spectral energy distribution (SED). GJ 876 is the coolest MUSCLES star with $\teff$=3129K and in our models the O$_2$ feature becomes most obscured for stars with $\teff$ = 2300K - 2600K.

\begin{figure}[h!]
\centering
\includegraphics[scale=0.34,angle=0]{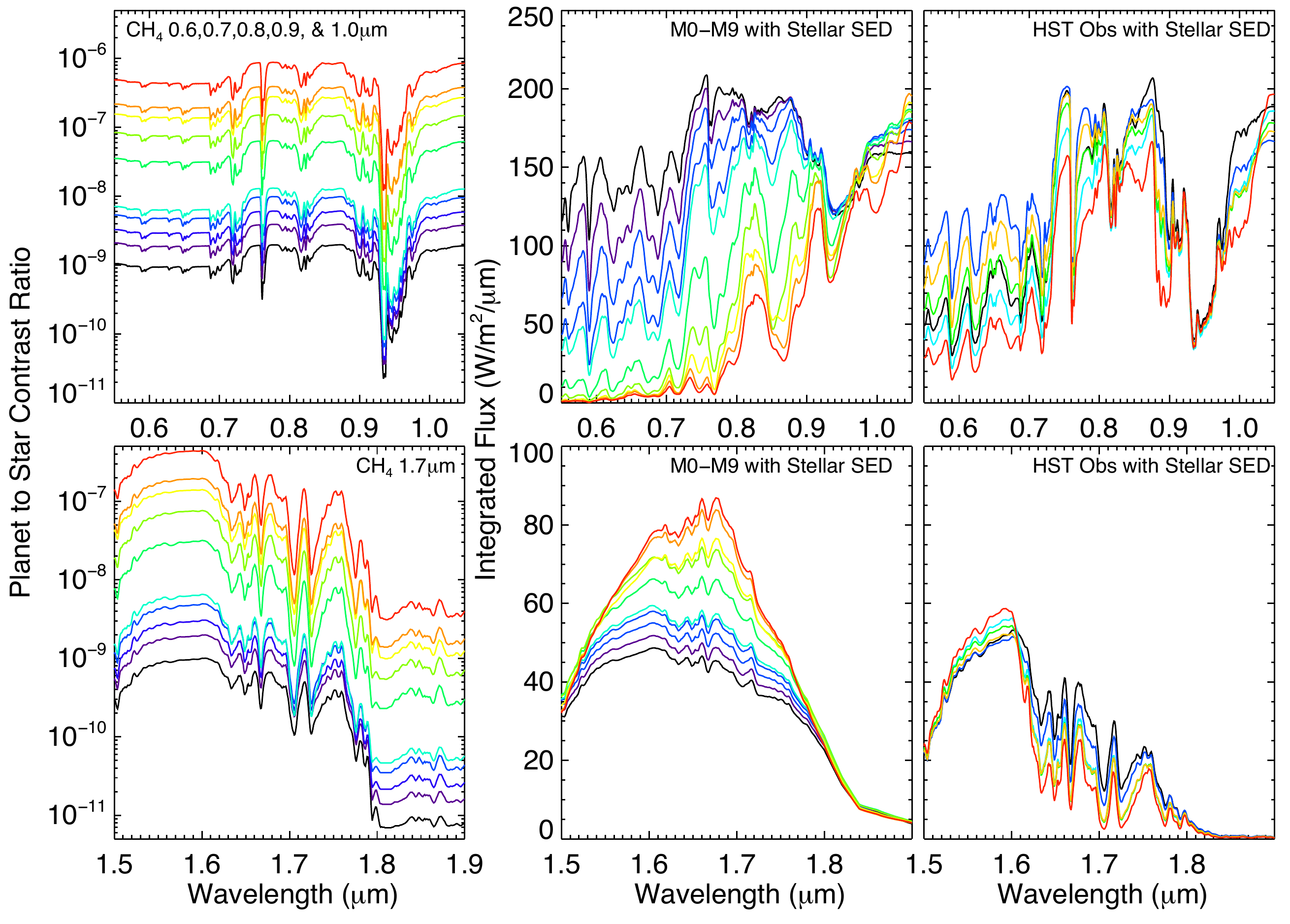}
\caption{CH$_4$ features at 0.6, 0.7, 0.8, 0.9, and 1.0$\mu$m (top row) and the CH$_4$ feature at 1.7$\mu$m (bottom row) for the relative flux as planet-to-star contrast ratio (left), and the reflected emergent flux for a 60\% cloud cover model for M0-M9 model stars (middle), and for MUSCLES stars (right). The observable CH$_4$ feature becomes deeper for planets orbiting cooler and less active star models. Coloring is same as in Fig. \ref{VISspectraMstars}.\label{methane}}
\end{figure}

CH$_4$ also has several features of interest in the VIS/NIR range at 0.6, 0.7, 0.8, 0.9, 1.0 and 1.7$\mu$m (see Fig. \ref{methane}). The CH$_4$ feature is deeper for less active and cooler M dwarfs. In particular, the CH$_4$ features at 0.8, 0.9 and 1.0$\mu$m become much more pronounced for the cooler M dwarfs and for some of the MUSCLES stars such as GJ 876. 

H$_2$O several features in the VIS/NIR at 0.95, 1.14 ,1.41 and 1.86$\mu$m and are deeper for cooler and less active stars (see Fig. \ref{VISspectraMstars}). 

Clouds increase reflectivity but can also decrease the depth of all observable features compared to the clear sky case. It will be difficult, therefore, to remotely determine the absolute abundance of a molecule without a well-characterized temperature vs. pressure distribution as well as cloud profile. 

\subsection{Earth-like Infrared Spectra, IR (4$\mu$m - 20$\mu$m)}

\begin{figure*}[ht!]
\centering
\includegraphics[scale=0.65]{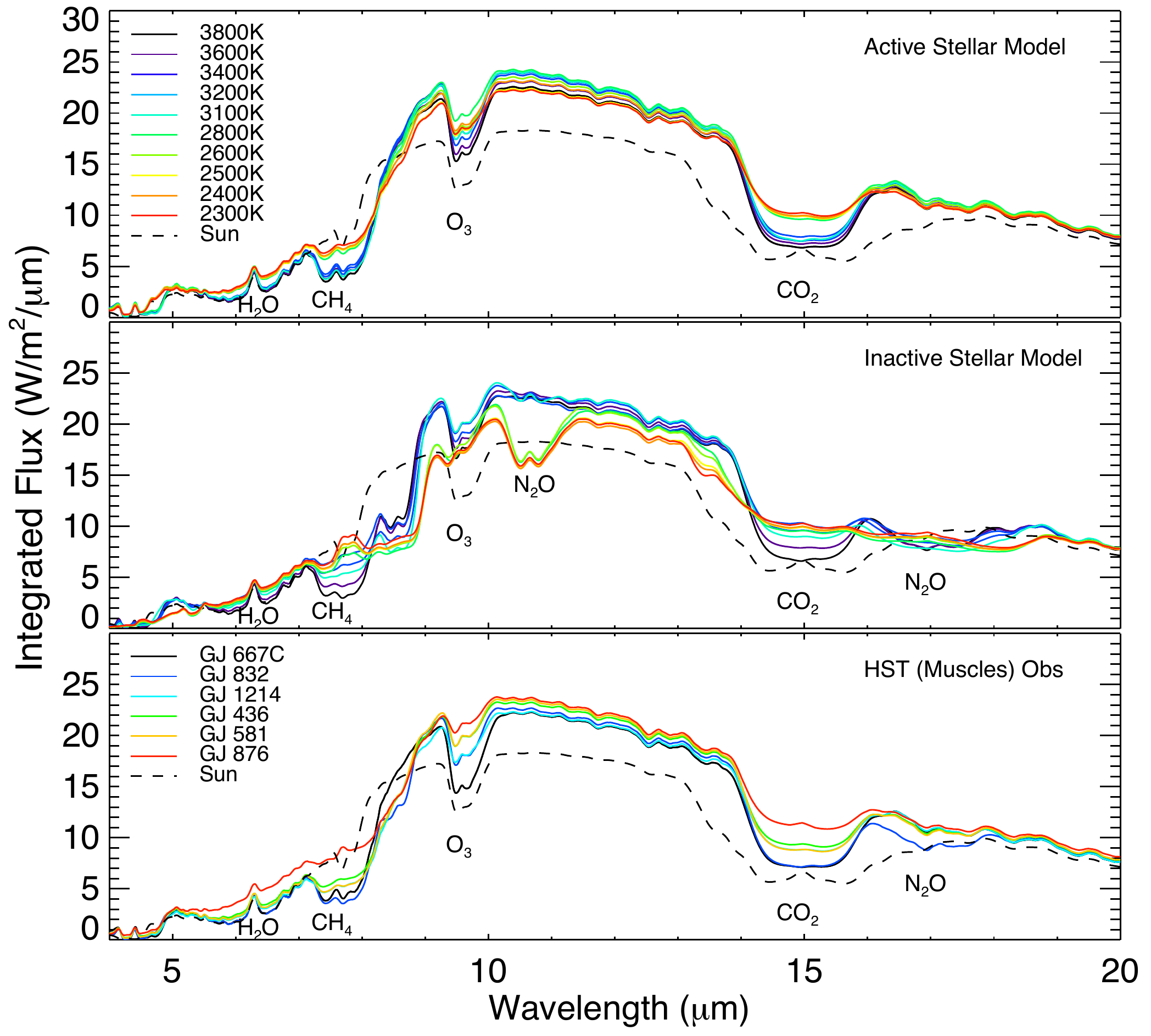}
\caption{Smoothed, disk-integrated IR spectra at the TOA emitted by an Earth-like planet orbiting the grid of active stellar models (top), inactive stellar models (middle), and MUSCLES stars (bottom) assuming 60\% Earth-analogue cloud coverage model. The Earth-Sun spectrum is shown for comparison as a dashed black line.\label{IRspectraMstars}}
\end{figure*}

\begin{figure*}[ht!]
\centering
\includegraphics[scale=0.55,angle=-90]{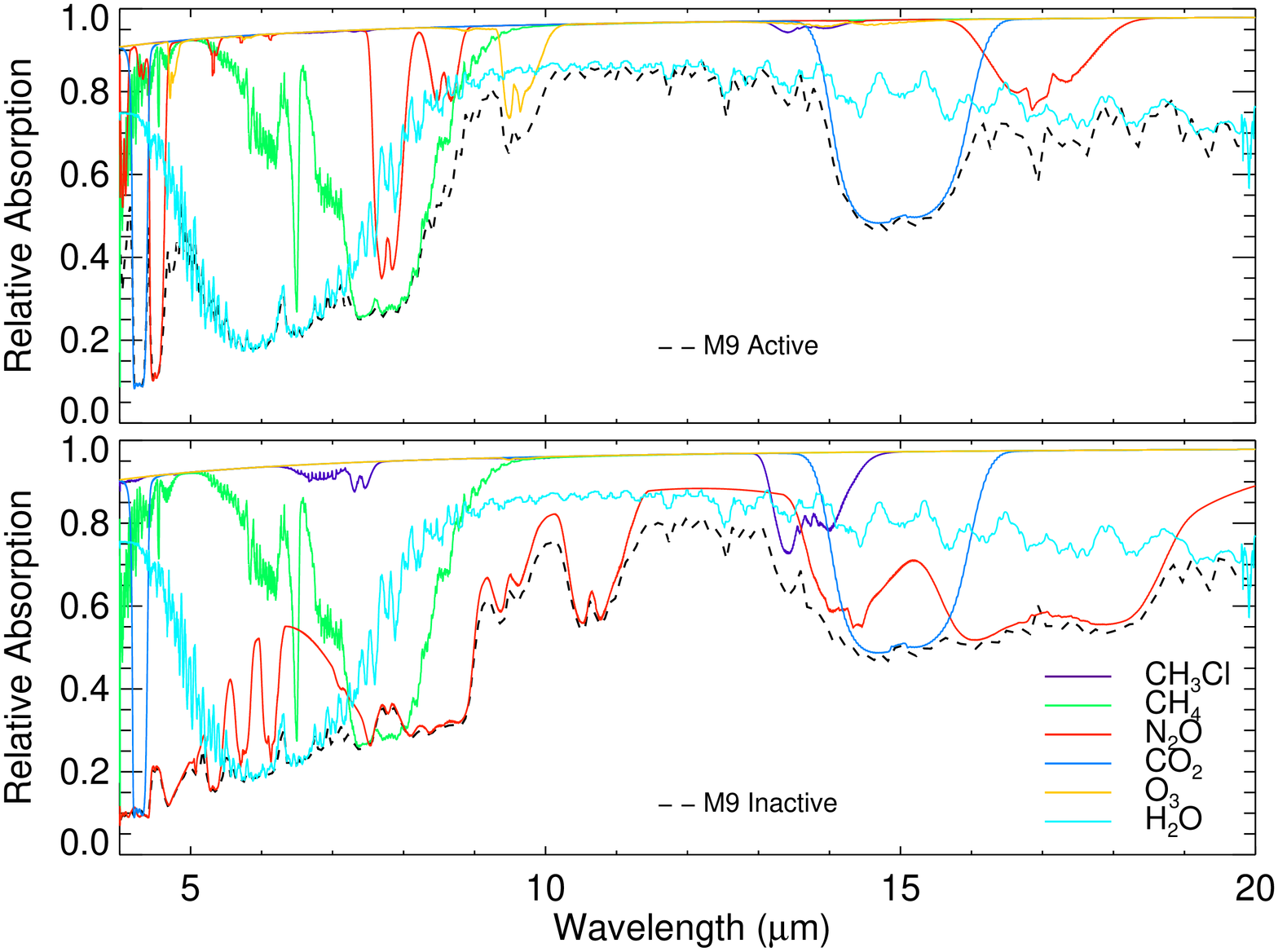}
\caption{Individual spectral components for H$_2$O, CH$_4$, CO$_2$, CH$_3$Cl, O$_3$, and N$_2$O comprising the full IR spectra (shown by a dashed black line) for an Earth-like planet orbiting a M9 active stellar model (top) and M9 inactive stellar model (bottom).\label{IRfeaturesMstars}}
\end{figure*}

Fig. \ref{IRspectraMstars} shows the thermal emission spectra from 4 to 20$\mu$m of Earth-like planets with Earth-analogue cloud cover around the M dwarf grid of active, inactive, and MUSCLES stars using the stellar SED. The high-resolution spectra have been smoothed to a resolving power of 150 using a triangular smoothing kernel to show the resolution expected by JWST. The Earth-Sun IR spectrum is shown for comparison as a dashed black line. Fig. \ref{IRfeaturesMstars} shows the individual component gas contributions of the dominate gases (H$_2$O, CH$_4$, CO$_2$, CH$_3$Cl, O$_3$, and N$_2$O) to the final IR planetary spectrum for an M9 active and inactive stellar model.

The depth of the O$_3$ feature at 9.6$\mu$m decreases for planets orbiting cooler and less active M dwarfs, as expected due to lower O$_3$ abundances for lower UV incident flux and also due to the decreased temperature difference between the O$_3$ emitting layer (around 40km) and the surface temperature, which is larger for the earlier M dwarfs as seen in Fig. \ref{tpchemMstars}.
 
The CH$_4$ feature at 7.7$\mu$m, decreases in depth for planets orbiting cooler M dwarfs despite increasing CH$_4$ abundances making it difficult to remotely determine the CH$_4$ abundance without a well-characterized temperature vs. pressure profile. Note that the 7.7$\mu$m feature is partially obscured by the wings of the H$_2$O feature at 5-8$\mu$m.

The N$_2$O features are the most striking addition in the inactive models with low UV flux. N$_2$O has features at 7.75$\mu$m (overlapping with the CH$_4$ feature), 8.5$\mu$m, 10.65$\mu$m and 16.9$\mu$m (see Fig. \ref{IRfeaturesMstars}) that become deeper with decreasing UV flux. In the active M0-M9 stellar models we do not see any strong N$_2$O features, although it does contribute to the depth of the 7.7 CH$_4$ feature. For the inactive M5-M9 models we see a strong N$_2$O feature at 10.65$\mu$m and wide absorption from 14-19$\mu$m due to the strong build-up of N$_2$O in low UV environments assuming an Earth-like biological surface flux of $7.99\times 10^{12}$ \mbox{g yr$^{-1}$} (see \S2 and \S3). However, even for modest mixing ratios of 15ppm to 34ppm in GJ 832 and the M0 inactive models, respectively, we see a strong N$_2$O feature at 16.9$\mu$m.  These concentrations are not much higher than the model mixing ratios calculated for the other MUSCLES stars and the later active M dwarf models. N$_2$O has been considered to be a strong biosignature \citep[see e.g.][]{segura2005, seager2012} and will be easier to detect around M dwarfs - especially inactive ones - than around FGK stars. 

The CO$_2$ absorption feature at 15$\mu$m does not have a central emission peak, usually seen in F and G stars, due to the more isothermal stratospheres of planets around all M dwarfs.

The depth of the H$_2$O feature at 5-8$\mu$m decreases for later M dwarfs despite increasing stratospheric H$_2$O concentrations. 

Clouds reduce the continuum level and the depth of all of the observable features compared to the clear sky case. Therefore, it will be difficult to remotely determine absolute abundances without a well-characterized temperature vs. pressure distribution as well as cloud profile.

\emph{Observability of Biosignatures}: Detecting the combination of O$_2$ or O$_3$ and a reducing gas like CH$_4$ in emergent spectra and in secondary eclipse measurements requires observations either in the IR between 7 and 10$\mu$m that includes the 7.7$\mu$m  CH$_4$ and 9.6$\mu$m O$_3$ features, or observations in the VIS to NIR between 0.7 to 3$\mu$m that includes the 0.76$\mu$m O$_2$ and 2.4$\mu$m CH$_4$ features in the observed spectral range. The strengths of absorption features (see Fig. \ref{VISspectraMstars} and \ref{IRfeaturesMstars} for VIS/NIR and IR, respectively) depend on the effective temperature of the host star as well as its UV flux and vary significantly between stellar types (see Fig. \ref{VISspectraMstars}-\ref{IRfeaturesMstars}). 

In the IR, CH$_4$ at 7.7$\mu$m is more easily detected at low resolution for earlier grid stars (M0-M4) than for later grid stars (M5-M9) in the active models. 

The 9.6$\mu$m O$_3$ feature is deepest in earlier and/or more active M dwarfs where there are more UV photons.  For M5-M9 inactive models, O$_3$ is not detectable in low resolution at 9.6$\mu$m due to the extremely low UV flux for these models. The narrow O$_2$ feature in the VIS at 0.72$\mu$m becomes weaker for cooler M dwarfs even though the mixing ratio of O$_2$ remains fixed in our simulations at 21\%. This is a consequence of the faint spectral energy distributions at shorter wavelengths for both active and inactive late M stellar models. 
   
For the modern Earth, N$_2$O and CH$_3$Cl do not contribute substantially to the spectrum due to their low mixing ratios and will likely be undetectable by the first low-resolution and photon-limited exoplanet atmosphere characterization missions \citep{selsis2000, kaltenegger2007}. However, both N$_2$O and CH$_3$Cl reach detectable levels in the IR spectra for our models of cooler and less active stars due to low photolysis rates even though N$_2$O has many absorption features in the IR (see Fig. \ref{IRfeaturesMstars}).

N$_2$O is considered a strong biosignature because there are no significant abiotic sources \citep{desmarais2002}.  One MUSCLES star, GJ 832, has a noticeable N$_2$O feature at 17$\mu$m with 15ppm for a modern Earth-like flux. This case is interesting because GJ 832 has the highest total FUV flux at wavelengths where N$_2$O is photolyzed (1000-2400\AA). We find that the high Ly-$\alpha$ flux of GJ 832 promotes the destruction of O$_3$, producing more O($^1$D) which then reacts with N$_2$ to form N$_2$O. 

For planets orbiting inactive M5-M9 star models we observe a sharp increase in N$_2$O concentrations as a consequence of few available UV photons. N$_2$O features dominate the spectrum for these late, inactive M dwarfs and can be detected in the IR at 4-5$\mu$m, 8-11$\mu$m, and 16-19$\mu$m (see Fig. \ref{IRfeaturesMstars}). In an atmosphere dominated by N$_2$O, such strong absorption throughout the IR could obscure the 7.7$\mu$m CH$_4$ feature. It is unknown whether a strong build-up of biotic N$_2$O would be physically possible around stars with little or no chromospheric flux. Given the existence of a few GALEX stars with little excess chromospheric NUV flux \citep{shkolnik2014} and only an upper limit established for the Ly-$\alpha$ flux from GJ 1214, a small fraction of M dwarfs could exhibit low UV fluxes. However, the MUSCLES data set shows that all six observed M dwarfs have sufficient UV flux to photolyze N$_2$O and to prevent ``run-away'' N$_2$O build-up. 

CH$_3$Cl contributes to our IR spectrum from 13-14$\mu$m in the short wavelength wing of the CO$_2$ and N$_2$O features and could also be detectable, depending on the CO$_2$ and N$_2$O concentrations present (see Fig. \ref{IRfeaturesMstars}).

For clear sky models, the vegetation red edge (VRE) surface feature is detectable in low resolution spectra due to the order of magnitude increased VRE reflectance between 0.7$\mu$m and 0.75$\mu$m for all M grid stars assuming that the exoplanets of these host stars have similar plant life \citep[see][for FGK stars]{rugheimer2013}. Clouds partly obscure this feature compared to the clear sky case, although the increase in flux can be seen at  0.7$\mu$m in Fig. \ref{VISspectraMstars} which include Earth-like clouds. Due to the shift in available photons to longer wavelengths for M dwarfs, a different photosynthesis biochemistry could have evolved, resulting in a different but potentially observable vegetation signature \citep{kiang2007}. 

\begin{figure*}[ht!]
\centering
\includegraphics[scale=0.6,angle=-90]{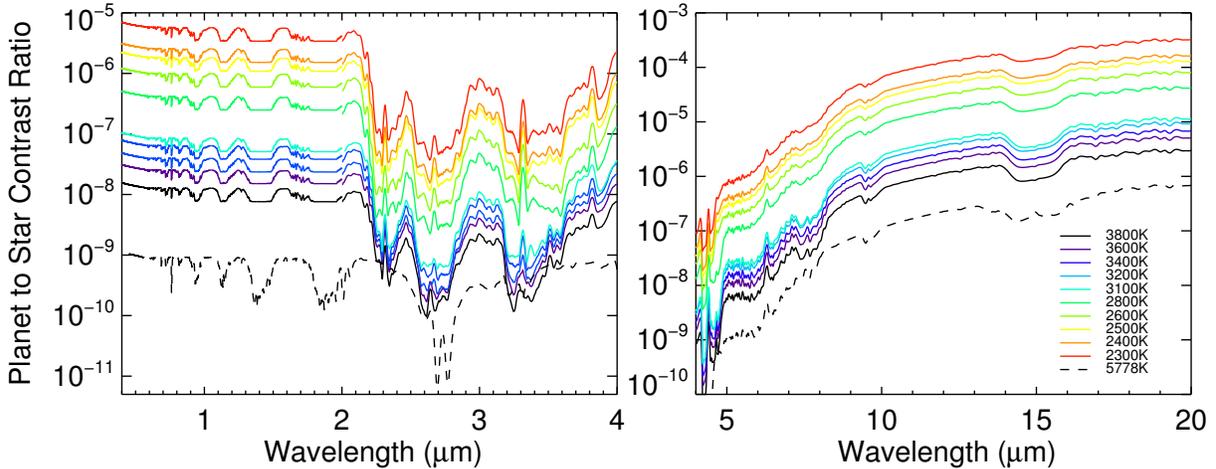}
\caption{Contrast ratios for Earth-like planets orbiting M0-M9 stars in full illumination. Active model contrast ratios are shown here. The contrast ratios of spectral features of planets orbiting inactive M dwarfs have similar levels as active M dwarfs in our models. The black dotted lines show contrast ratios for the Earth orbiting the Sun.\label{contrastMstars}}
\end{figure*}

\emph{Observability of Spectral Features}
Note that we have not added noise to these model spectra in order to be useful as input models for a wide variety of instrument simulators for both secondary eclipse and direct detection simulations. Different instrument simulators for JWST \citep[see e.g.][]{deming2009, kaltenegger2009} explore the capability of JWST's MIRI and NIRspec instruments to characterize extrasolar Earth-like planets for near-by as well as luminous host stars. Several groups are providing realistic instrument simulators that can be used to determine the detectability of these absorption features. Future ground and space based telescopes are being designed to characterize exoplanets as small as Earth-like planets and will provide opportunities to observe atmospheric features, especially for Super-Earths with radii up to twice Earth's radius and therefore four times the flux and planet-to-star contrast ratio levels for Earth-size planets as shown in Fig. \ref{contrastMstars}. 

In addition to measuring the size of the planet, future observations will occur at different phases throughout the planet's orbit. The maximum observable planetary flux in the visible spectrum scales with the illuminated fraction of the planet that is visible to the observer. At quadrature, representing an average viewing geometry, the contrast ratios presented in Fig. \ref{contrastMstars} will be a factor of $\sim$ 2 lower in the visible. In the IR, the maximum flux remains constant throughout the planet's orbit, assuming a similar temperature on the planet's day and night side.

\section{DISCUSSION}

Our active and inactive stellar models represent the extreme ends of stellar activity for M dwarfs. AD Leo, our active star proxy with H$_\alpha$ in emission, was chosen to represent a young, active star. Our inactive stellar models assume no chromospheric activity, which is the lowest level possible of UV flux. More observations are needed to determine the lowest possible level of UV flux around an M dwarf, which still might be significantly higher than the photosphere only models. Many M dwarfs exhibit a wide range of activity. Therefore, we use the MUSCLES star database to represent observed M dwarfs with a range of activity between the two extremes. The simulated planets orbiting the MUSCLES stars share more similarities with the planets orbiting the active stellar models particularly in terms of O$_3$, CH$_4$, and N$_2$O concentrations, which is to be expected since they have UV fluxes detectable with HST.

Ly-$\alpha$ is an important line to characterize because it is the strongest line in the FUV part of an M dwarf spectrum. We note that in addition to properly reconstructing the Ly-$\alpha$ line, determining the base level of flux and other emission lines in the FUV and NUV are also important. This is because Ly-$\alpha$ by itself does not significantly impact the photochemistry of an Earth-like planet atmosphere below about 60km unless the  Ly-$\alpha$ flux exceeds the highest observed levels observed for the MUSCLES stars (see \S3 and Fig. \ref{tpchemlya}).

Our present models assume a steady state. Young M dwarfs exhibit strong flares which may impact the atmosphere on timescales relevant to the photochemistry. Future work will consider activity induced variability in an exoplanet's atmosphere. See \citet{segura2010} for models on time-dependent behavior of biosignatures for flaring M dwarfs.

\section{CONCLUSIONS}

We show  spectral features for terrestrial atmosphere models of Earth-like planets from the VIS to the IR for planets orbiting a grid of M0 to M9 host stars ($\teff$ = 2300K to 3800K) using active stellar models, inactive stellar models, and the sample of M dwarfs with observed UV fluxes. Our grid comprehensively covers the full M stellar range and activity from a proxy of an active flare star, AD Leo, to the lower limit of activity with no chromospheric contribution to the UV. 

We discuss the atmospheric model results in \S3 and the detectable spectral features with a focus on how the UV flux environment affects the emergent spectra in \S4. Increased UV flux environments in M dwarfs is primarily a property of younger and earlier type stars. Higher UV environments produce: increased concentration of O$_3$ from photolysis of O$_2$, increased O$(^1D)$ from photolysis of O$_3$, increased OH from reaction of O$(^1D)$ and H$_2$O, decreased stratospheric H$_2$O, and decreased CH$_4$, CH$_3$Cl, and N$_2$O from photolysis and reactions with OH (see Fig. \ref{tpchemMstars}). Very inactive M dwarfs may have extremely low UV flux levels as suggested by GALEX observations in the NUV. More observations are needed to determine the lower limit of fluxes in the FUV, in particular the Ly-$\alpha$ emission. Such low levels of UV flux could lead to a build-up molecules such as CH$_4$ and N$_2$O in the atmosphere of an Earth-like planet, with N$_2$O being a strong biosignature due to a lack of photochemical sources.

Terrestrial planets orbiting early M spectral type stars are the best targets for observing biosignatures, such as O$_2$ or O$_3$ in combination with a reducing species like CH$_4$. Note that the O$_2$ line becomes increasingly difficult to detect for late M dwarfs (see Fig. \ref{oxygenmstars}). While not observable for an exoplanet around an FGK star, N$_2$O may be observable at 10.7$\mu$m or at 17$\mu$m at Earth-level emissions for planets orbiting M dwarfs (see inactive models and model for GJ 832). 
  
Our results provide a grid of atmospheric compositions as well as model spectra from the VIS to the IR for JWST and other future space and ground-based direct imaging and secondary eclipse missions in terms of instrument design and will help to optimize their observation strategy. The model spectra in this paper are available at www.cfa.harvard.edu/$\sim$srugheimer/Mspectra/.

\section*{Acknowledgments}

We especially thank Andrew West and Evgenya L. Shkolnik for useful conversations about M dwarfs. We also would like to thank Kevin France for discussions concerning the MUSCLES database. This work has made use of the MUSCLES M dwarf UV radiation field database. We would also like to acknowledge support from DFG funding ENP KA 3142/1-1 and the Simons Foundation (290357, Kaltenegger).

The UV radiation field for AD Leo was obtained from the Multimission Archive at the Space Telescope Science Institute (MAST). STScI is operated by the Association of Universities for Research in Astronomy, Inc., under NASA contract NAS5-26555. Support for MAST for non-HST data is provided by the NASA Office of Space Science via grant NAG5-7584 and by other grants and contracts.

\section*{Appendix}

\textbf{GJ 667C}: GJ 667C is an M3-4 dwarf at 6.8 pc \citep{vanleeuwen2007} with an age estimate greater than 2 Gyr \citep{anglada2013b}. GJ 667C has a $\teff$ = 3350K, [Fe/H] = -0.55, and M = 0.330$M_{\odot}$ $\pm$ 0.019 \citep{anglada2013b}. Previous estimates had the effective temperature 350K higher and the spectral designation to be M1.5 due to assuming a higher metallicity \citep[see][]{geballe2002, anglada2013b}. We use a radius of R = 0.35$R_{\odot}$ consistent with the luminosity, L = 0.01370 $L_{\odot}$ \citep{anglada2012}. We merged a PHOENIX BT-Settl spectrum for $\teff$ = 3350K, [Fe/H] = -0.55, log(g) = 5, and $v$ sin $i$ = 3 km s$^{-1}$ with the MUSCLES UV spectrum at 2800 \AA. GJ 667C has less wavelength coverage than the other MUSCLES stars. We approximated the rest of UV radiation field from the FUV and NUV spectral energy distribution of GJ 832, which has a similar spectral type \citep{france2013}, scaled to the distance and Ly-$\alpha$ and Mg II emission line strength of GJ 667C.

\textbf{GJ 832}: GJ 832 is an M dwarf at 4.95 pc \citep{vanleeuwen2007} with no age estimate and is the least well characterized star in the MUSCLES program sample. It has a $\teff$ = 3620K \citep[NStED value interpolated as described in][]{bessell1995}, R = 0.48$R_{\odot}$  \citep{johnson1983}, [Fe/H] = -0.12 \citep{johnson2009}, and log $g$ = 4.7 \citep{bailey2008}. We merged a PHOENIX BT-Settl spectrum with $\teff$ = 3620K, [Fe/H] = -0.12, log $g$ = 4.7, and a $v$ sin $i$ = 3 km s$^{-1}$ with the HST UV spectra from the MUSLES program at 2800 \AA.

\textbf{GJ 1214}: GJ 1214 is an M6 dwarf at 14.55$\pm$ 0.13pc \citep{anglada2013a} with an age estimate of 6$\pm$3 Gyr \citep{charbonneau2009}. A more accurate parallax measurement increased the previous distance estimate by 10\% and thus the luminosity and mass has been shifted from previous values as well \citep{anglada2013a}. With the new parallax and luminosity, GJ 1214 has a radius of 0.211 $R_\odot$, $\teff$ = 3250K, M = 0.176 $M_\odot$ \citep{anglada2013a}, and [Fe/H] = +0.05 \citep{anglada2013a, neves2012}. We used an upper limit of $v$ sin $i$ = 1 km s$^{-1}$ \citep{browning2010, delfosse1998, reiners2008,west2009} and log $g$ = 4.991 \citep{charbonneau2009}. We merged a PHOENIX BT-Settl spectrum with $\teff$ = 3250K, [Fe/H] = +0.05, log $g$ = 4.991, and $v$ sin $i$  = 1 km s$^{-1}$ with the MUSCLES UV spectrum at 2800 \AA. GJ 1214 is the only MUSCLES star to have a non-detection of Ly-$\alpha$, and thus an upper limit is used and discussed in depth in \citet{france2013}. 

\textbf{GJ 436}: GJ 436 is an M3 dwarf at 10.1 pc \citep{vanleeuwen2007} with an age estimate of 6.5-9.9 Gyr \citep{saffe2005}. GJ 436 has a $\teff$ = 3416K, R = 0.455$\pm$ 0.018 $R_\odot$, and M = 0.507 (+0.071/-0.062) $M_\odot$ \citep{vonbraun2012}. GJ 436 has solar metallicity, [Fe/H]=0 \citep{rojas2010}, log $g$ = 5.0 \citep{maness2007} and a $v$ sin $i <$ 1 km s$^{-1}$ \citep{marcy1992}. We merged a PHOENIX BT-Settl spectrum with $\teff$ = 3416K, [Fe/H] = 0, log $g$ = 5 and a $v$ sin $i$ = 1 km s$^{-1}$ with the HST UV spectra from the MUSCLES program at 2800 \AA.

\textbf{Gl 581}: GJ 581 is a M3 dwarf at 6.2$\pm$ 0.1pc \citep{vanleeuwen2007} with age estimate of 7-11  Gyr \citep{selsis2007}. GJ 581 has a $\teff$ = 3498.0K$\pm$ 56.0K, R = 0.299 $\pm$ 0.010 $R_\odot$, and log $g$ = 4.96$\pm$0.08 \citep{vonbraun2011}. GJ 581 has a metallicity slightly subsolar, [Fe/H] = -0.02 \citep{rojas2010} and an upper limit on $v$ sin $i \le \ $2.1 km s$^{-1}$ \citep{delfosse1998}. We merged a PHOENIX BT-Settl spectrum with $\teff$ = 3498K, [Fe/H] = -0.02, log $g$ = 4.96, and $v$ sin $i$ =  2.1 km s$^{-1}$ \citep{correia2010} with the MUSCLES UV spectrum at 2800 \AA.

\textbf{GJ 876}: GJ 876 is an M dwarf at 4.69pc \citep{vanleeuwen2007} with an age 0.1-5Gyr \citep{correia2010}. GJ 876 has $\teff$ = 3129$\pm$19K and R = 0.3761 $\pm$ 0.0059 $R_{\odot}$  \citep{vonbraun2014}. We merged a PHOENIX BT-Settl spectrum with $\teff$ = 3129K, [Fe/H] = +0.19 \citep{rojas2012}, log $g$ = 4.89 \citep{bean2006}, and $v$ sin $i$ = 1.38 km s$^{-1}$ \citep{correia2010} with the MUSCLES UV spectrum at 2800 \AA.

\end{document}